\documentclass[fleqn,usenatbib]{mnras}
\usepackage{newtxtext,newtxmath}

\usepackage[T1]{fontenc}

\DeclareRobustCommand{\VAN}[3]{#2}
\let\VANthebibliography\thebibliography
\def\thebibliography{\DeclareRobustCommand{\VAN}[3]{##3}\VANthebibliography}

\usepackage{graphicx}	
\usepackage{amsmath}	
\usepackage{multirow}
\usepackage{booktabs}
\usepackage{enumitem}
\usepackage{tikz}
\usepackage{array}
\usepackage[normalem]{ulem}
\usepackage{soul}
\usepackage{rotating}
\usepackage{graphicx}
\usepackage{pdflscape}
\newcolumntype{L}[1]{>{\raggedright\let\newline\\\arraybackslash\hspace{0pt}}m{#1}}
\newcolumntype{C}[1]{>{\centering\let\newline\\\arraybackslash\hspace{0pt}}m{#1}}
\newcolumntype{R}[1]{>{\raggedleft\let\newline\\\arraybackslash\hspace{0pt}}m{#1}}

\title[UVIT catalogue and membership of OCs]{UOCS\thanks{UVIT open cluster study}. III. UVIT catalogue of open clusters with machine learning based membership using \textit{Gaia} EDR3 astrometry}

\author[Jadhav et al.]{
Vikrant V. Jadhav$^{1,2}$\thanks{E-mail: vikrant.jadhav (at)iiap.res.in},
Clara M. Pennock$^{3}$,
Annapurni Subramaniam$^{1}$,
Ram Sagar$^{1}$,
\newauthor
and Prasanta Kumar Nayak$^{4}$
\\
$^{1}$Indian Institute of Astrophysics, Koramangala II Block, Bangalore-560034, India\\
$^{2}$Joint Astronomy Program and Physics Department, Indian Institute of Science, Bangalore-560012, India\\
$^{3}$Lennard-Jones Laboratories, Keele University, Staffordshire ST5 5BG, UK\\
$^{4}$Tata Institute of Fundamental Research, Colaba, Mumbai-400005, India
}

\date{Accepted XXX. Received YYY; in original form ZZZ}

\pubyear{2020}

\begin{document}
\label{firstpage}
\pagerange{\pageref{firstpage}--\pageref{lastpage}}
\maketitle

\begin{abstract}
We present a study of six open clusters (Berkeley 67, King 2, NGC 2420, NGC 2477, NGC 2682 and NGC 6940) using the Ultra Violet Imaging Telescope (UVIT) aboard \textit{ASTROSAT} and \textit{Gaia} EDR3. We used combinations of astrometric, photometric and systematic parameters to train and supervise a machine learning algorithm along with a Gaussian mixture model for the determination of cluster membership. This technique is robust, reproducible and versatile in various cluster environments. In this study, the \textit{Gaia} EDR3 membership catalogues are provided along with classification of the stars as \texttt{members, candidates} and \texttt{field} in the six clusters. We could detect 200--2500 additional members using our method with respect to previous studies, which helped estimate mean space velocities, distances, number of members and core radii. UVIT photometric catalogues, which include blue stragglers, main-sequence and red giants are also provided. From UV--Optical colour-magnitude diagrams, we found that majority of the sources in NGC 2682 and a few in NGC 2420, NGC 2477 and NGC 6940 showed excess UV flux. NGC 2682 images have ten white dwarf detection in far-UV. The far-UV and near-UV images of the massive cluster NGC 2477 have 92 and 576 \texttt{members} respectively, which will be useful to study the UV properties of stars in the extended turn-off and in various evolutionary stages from main-sequence to red clump. Future studies will carry out panchromatic and spectroscopic analysis of noteworthy members detected in this study.
\end{abstract}




\section{Introduction} \label{sec:intro}

Star clusters are test-beds for the study of stellar evolution of single, as well as, binary stars in diverse physical environments. Multi-wavelength studies of stars in clusters help to reveal the possible formation mechanism of non-standard stellar populations \citep{Thomson2012, Jadhav2019}. Open clusters (OCs) in Milky Way span a wide range in ages, distances and chemical compositions \citep{Dias2002, Kharchenko2013, Netopil2016, Cantat2020}. The relatively low stellar density in the OCs is also an essential factor which helps in understanding properties of binary systems in a tidally non-disruptive environment. 

\begin{table*}
    \centering
    \caption{Cluster coordinates, ages, distances, mean PMs and radii are taken from \citet{Cantat2018,Cantat2020}. The metallicity of Berkeley 67 is from \citet{Lata2004} and other metallicities are from \citet{Dias2002}.}
    \label{tab:cluster_details}
    \begin{tabular}{lcccc ccccc r}
    \toprule
Name	&	$\alpha_c$ (J2015.5)	&	$\delta_c$ (J2015.5)	&	l	&	b	&	$D$	&	Age	&	[M/H]	&	$\mu_{\alpha, c}cos\delta$	&	$\mu_{\delta, c}$	&	r50	\\
	&	($^{\circ}$)	&	($^{\circ}$)	&	($^{\circ}$)	&	($^{\circ}$)	&	(pc)	&	(Gyr)	&		&	(mas yr$^{-1}$)	&	(mas yr$^{-1}$)	&	(')	\\ \hline
Berkeley 67	&	69.472	&	50.755	&	154.85	&	2.48	&	2216	&	1.3	&	+0.02	&	2.3	&	-1.4	&	4.9	\\
King 2	&	12.741	&	58.188	&	122.87	&	-4.68	&	6760	&	4.1	&	-0.41	&	-1.4	&	-0.8	&	3.1	\\
NGC 2420       	&	114.602	&	21.575	&	198.11	&	19.64	&	2587	&	1.7	&	-0.38	&	-1.2	&	-2.1	&	3.2	\\
NGC 2477	&	118.046	&	-38.537	&	253.57	&	-5.84	&	1442	&	1.1	&	+0.07	&	-2.4	&	0.9	&	9.0	\\
NGC 2682       	&	132.846	&	11.814	&	215.69	&	31.92	&	889	&	4.3	&	+0.03	&	-11.0	&	-3.0	&	10.0	\\
NGC 6940	&	308.626	&	28.278	&	69.87	&	-7.16	&	1101	&	1.3	&	+0.01	&	-2.0	&	-9.4	&	15.0	\\
   \toprule
    \end{tabular}
\end{table*} 

The OCs of our Galaxy are located at various distances from us. Thus, stars detected in any observation will be a mixture of cluster members as well as both foreground and background field stars. The identification of cluster members using a reliable method is therefore extremely important. Earlier this was accomplished using the spatial location of stars in the cluster region, as well as, their location on different phases of a single star evolution i.e. the main sequence, sub-giant branch and red giant branch in the colour magnitude diagrams (CMDs) of star clusters \citep{Shapley1916}. 
However, many intriguing and astrophysically significant stars such as blue straggler stars (BSS), sub-sub-giants were not considered members due to their \textit{peculiar} locations in the OC CMDs (we refer to locations other than main sequence, sub-giant branch and red giant branch, which are part of the single star evolution, as \textit{peculiar}). In order to study and understand progenitors of these stars, it is vital to establish their cluster membership. More importantly, we also should be able to estimate the membership probability of these \textit{peculiar} stars, which will bring confidence when they are considered as members of an OC.

Using the proper motion (PM) of stars in the vicinity of the clusters, \citet{van1942} made significant improvements in the membership determination.
\citet{Vasilevskis1958} and \citet{Sanders1971} pioneered the techniques of membership probability (MP) determination using vector point diagrams (VPDs). As the accuracy of PM measurements improved, membership determination using VPDs were also enhanced (\citealp{Sagar1987, Zhao1990, Balaguer1998, Bellini2009} and references therein). Assuming that the field and cluster stars produce overlapping Gaussian distributions in the VPD, new techniques were developed to separate the two populations \citep{Bovy2011, Vasiliev2019}.
The arrival of \textit{Gaia} was instrumental in study of star clusters \citep{Gaia2016}. Trigonometric parallax and accurate PM from \textit{Gaia} DR2 has led to accurate identification of cluster members: Hertzsprung--Russell diagram of star clusters help model the stellar evolution \citep{Gaia2018a}; \citet{Cantat2018} identified cluster members in 1229 OCs and discovered 60 new clusters using ($\mu_{\alpha}, \mu_{delta}, \varpi$) clustering. Similarly, \citet{Liu2019, Sim2019, He2020, Castro2020} discovered new OCs by applying visual or machine learning techniques to \textit{Gaia} DR2 and identified cluster members.

We started a long-term program 
for the study of UV bright stellar population in the OCs using Ultra Violet Imaging Telescope (UVIT) payload mounted on the \textit{ASTROSAT}; the first Indian multi-wavelength space observatory launched successfully on 28th September 2015. Our initial studies were centred around the well-studied OCs, NGC 2682 (M67) and NGC 188 \citep{Subramaniam2016, Jadhav2019}. In NGC 2682, we found many chromospherically active stars, single and massive white dwarfs (WDs), and post-mass transfer systems (BSSs, BSS+WD, etc.), particularly those with extremely low mass WDs. Their detection and characterisation are necessary to study their UV energy budget, emission mechanism and formation pathways.

\begin{table*}
    \centering
    \caption{The log of UVIT observations in different filters are given along with exposure time and FWHM. The number of detected stars in each filter as well as the number of cluster \texttt{members/candidates} determined according to the Eq.~\ref{eq:classif} are also listed.}
    \label{tab:cluster_obs}
    \begin{tabular}{lccc cccr} 
    \toprule
Cluster	&	Filter	&	Observation Date	&	Exposure Time	&	Detected Stars	&	\texttt{Members}	&	\texttt{Candidates}	&	FWHM	\\	
	&		&	(yyyy-mm-dd)	&	(s)	&		&		&		&	(\arcsec)	\\	\toprule
Berkeley 67	&	N242W	&	2016-12-21	&	2700	&	469	&	64	&	5	&	1.15	\\	
	&	N245M	&	2016-12-21	&	2722	&	258	&	19	&	3	&	1.09	\\	\hline
King 2	&	F148W	&	2016-12-17	&	2666	&	150	&	5	&	1	&	1.33	\\	
	&	N219M	&	2016-12-17	&	2714	&	303	&	3	&	1	&	1.35	\\	\hline
NGC 2420	&	F148W	&	2018-04-30	&	2136	&	177	&	57	&	2	&	1.70	\\	\hline
NGC 2477	&	F148W	&	2017-12-18	&	2278	&	301	&	92	&	16	&	1.56	\\	
	&	N263M	&	2017-12-18	&	1881	&	1637	&	576	&	53	&	1.34	\\	\hline
NGC 2682	&	F148W	&	2018-12-19	&	6575	&	918	&	84	&	18	&	1.76	\\	
	&	F154W	&	2017-04-23	&	2428	&	267	&	31	&	7	&	1.47	\\	
	&	F169M	&	2018-12-19	&	6596	&	259	&	58	&	15	&	2.01	\\	\hline
NGC 6940	&	F169M	&	2018-06-13	&	1875	&	151	&	43	&	12	&	1.73	\\	
\toprule
    \end{tabular}
\end{table*} 

In order to identify such types of stars in other OCs, 
we selected clusters which are safe to be observed using UVIT (those at high galactic latitude and without bright stars in the UVIT field of view) and also have high probability of detecting UV stars. In this work, we extend our UVIT study of OCs to Berkeley 67, King 2, NGC 2420, NGC 2477, NGC 2682 and NGC 6940. They span a range of age (0.7 to 6 Gyr) and distance (0.8 to 5.8 kpc). Table~\ref{tab:cluster_details} lists the parameters such as location in the sky, distance, age, mean PM and radius of the OCs under study. The relevant literature surveys are included in Appendix \ref{sec:intro_1}.

We presented the results from NGC 2682 observations in \citet{Sindhu2019} and \citet{Jadhav2019}. In this paper, we include more recent and deeper photometry for NGC 2682.
It is also one of the most studied OCs with well-established CMD; hence we compare the behaviour of other OCs with NGC 2682 to interpret the optical and UV CMDs in further sections. We also use it to validate our membership determination method against previous efforts.
This paper aims to analyse the UV--optical CMDs and overall UV characteristics of these clusters. A detailed study of the noteworthy sources will be presented separately.

We used multi-modal astrometric and photometric data from latest \textit{Gaia} EDR3 \citep{Gaia2020_summary} to create a homogeneous catalogue of cluster members in the OCs. The membership determination of OC stars, in particular the UV bright population of BSSs, binaries and WDs, requires careful incorporation data quality indicators from \textit{Gaia} EDR3. The reduced errors in \textit{Gaia} EDR3 ($\sim$ 0.71x for parallax and $\sim$0.44x for PM; \citealt{Lindegren2020,Riello2020}) has therefore been used in the membership selection process. 
PMs of field and cluster stars can be approximated by Gaussian distributions \citep{Sanders1971} which can be separated analytically, and individual MP can be estimated from the distance of a star from field and cluster centre in the VPD. However, this method does not distinguish between field stars which have the same PM as cluster members. 
Therefore, parallax and CMD position could be used to remove field stars.
Also, parallax, colour and magnitudes have non-Gaussian distributions. In order to optimally use all the \textit{Gaia} parameters,
we chose supervised machine learning (ML) to segregate the cluster members. Use of ML techniques is increasing in astronomy to automate classification tasks, including cluster membership \citep{Gao2018a, Gao2018b, Gao2018c, Gao2018d, Zhang2020, Castro2020}.
However, as most ML techniques do not include errors in the data, we used probabilistic random forest ({\sc prf}, \citealt{Reis2019}), which incorporates errors in the data. To train the {\sc prf}, we first selected the cluster members by deconvolving the PM Gaussian distributions using a Gaussian Mixture Model ({\sc gmm}, \citealt{Vasiliev2019}). The overall method also provides the much-needed MPs, which are necessary for stars which follow the non-standard evolution.

\begin{table*}
\centering
\caption{Definition and formulae for \textit{Gaia} parameters and derived parameters used in this study.}
\label{tab:Gaia parameters}
\begin{tabular}{C{15cm}} 	\toprule					
\textbf{\textit{Gaia} EDR3 Parameters $^{\dagger}$} 	\\ \midrule					
\textsc{ra, dec, pmra, pmdec, parallax, ra\_error, dec\_error, pmra\_error, pmdec\_error, parallax\_error,} \\ 
{\sc phot\_g\_mean\_mag (\text{AS} g), phot\_bp\_mean\_mag (\text{AS} gbp), phot\_rp\_mean\_mag (\text{AS} grp), bp\_rp, bp\_g, g\_rp, phot\_g\_mean\_flux, phot\_bp\_mean\_flux, phot\_rp\_mean\_flux, phot\_g\_mean\_flux\_error, phot\_bp\_mean\_flux\_error, phot\_rp\_mean\_flux\_error, g\_zero\_point\_error, bp\_zero\_point\_error, rp\_zero\_point\_error, } \\ 
\textsc{astrometric\_excess\_noise (\text{AS} aen), astrometric\_excess\_noise\_sig (\text{AS} aen\_sig), ruwe}	\\ \toprule	
\textbf{Cluster parameters}  (taken from Table~\ref{tab:cluster_details})
	\\ \midrule					
\textit{ra\_cluster, dec\_cluster, parallax\_cluster, radius\_cluster}	\\ \toprule					\textbf{Derived parameters}	\\ \midrule	
\end{tabular}						
\begin{tabular}{L{3cm}L{8cm}R{3cm}}						
mag\_error	&	$\sqrt{\left(1.086 \times \frac{flux\_error}{flux}\right)^2 + zero\_point\_error^2}$	&	[mag]	\\	
\textsc{pmR0}	&	$\sqrt{(\textsc{pmra}-cluster\_pmra)^2+(\textsc{pmdec}-cluster\_pmdec)^2}$	&	[mas yr$^{-1}$]	\\	
\multirow{2}{*}{$\textsc{qf}\  \text{ (Quality filter)} \quad \quad \bigg\{ $}	&	0, $\qquad \text{if } \textsc{(ruwe > 1.4) \quad or \quad (aen > 1 and aen\_sig > 2)}$	&		\\	
	&	$1, \qquad \text{otherwise}$	&		\\ \toprule	
\end{tabular}
\\$^{\dagger}$ {\footnotesize \url{gea.esac.esa.int/archive/documentation/GEDR3/Gaia_archive/chap_datamodel/sec_dm_main_tables/ssec_dm_gaia_source.html}}
\end{table*}  

The paper is arranged as follows: section~\ref{sec:obs} has the details of UVIT observations, \textit{Gaia} data and isochrone models. 
The membership determination technique is explained in section~\ref{sec:method}.
The membership results and UV--optical photometry are presented in section~\ref{sec:results} and discussed in section~\ref{sec:discussion}. Supplementary tables and figures are included in the Appendix.
The full versions of \textit{Gaia} EDR3 membership catalogue (Table~\ref{tab:cat_Gaia}) and UV photometric catalogues of the six OCs (Table~\ref{tab:cat_UV}) are available online.

\section{Data and Models} \label{sec:obs}

\subsection{UVIT data} \label{dec:uvit_data}

The observations were carried out during December 2016 to December 2018 using different UV filters of \textit{ASTROSAT} UVIT payload (\textit{ASTROSAT} proposal IDs: A02\_170, A04\_075, G07\_007 and A05\_068). The log of UVIT observations is presented in Table~\ref{tab:cluster_obs}, along with total exposure time in each filter. We planned observations in at least one FUV and one NUV broadband filter in order to get wavelength coverage across the UV regime for detailed study.
Due to payload related issue, NUV observations were done only for early observations such as Berkeley 67, King 2 and NGC 2477.
Unfortunately, no cluster members could be detected in UV observations of Berkeley 67 despite observing it in two FUV filters due to lack of FUV bright stars.
The remaining three OCs (NGC 2420, NGC 2682 and NGC 6940), were observed in FUV filters alone.

Exposure time for different filters ranges from 1875 sec to 6596 sec with a typical value of $\sim$ 2000 sec. We used {\sc ccdlab} to create science ready images by correcting for drift, field distortions, and flat fielding \citep{Postma2017}. Astrometry was done by cross-matching stars from UVIT (point spread function, PSF $\sim 1.^{\prime\prime}5$) and \textit{Gaia} DR2 data (PSF $\sim 0.^{\prime\prime}4$) using {\sc iraf}. The PSF full width at half maximum (FWHM) of UVIT observations ranges from 1.$^{\prime\prime}$1 to 1.$^{\prime\prime}$8 with a mean value of $\sim$ 1.$^{\prime\prime}$4. The circular field of view of the UVIT has a radius of 14$\arcmin$, whereas the typical radius of the usable UVIT images is slightly less \citep{Tandon2017a}. 

We performed PSF photometry on all UVIT images using {\sc daophot} package of {\sc iraf}. We used 5-$\sigma$ detection and limited the catalogue up to the detection with magnitude errors $<$ 0.4 mag. The magnitude vs PSF error plots for all the images are shown in Fig.~\ref{fig:mag_err}. The magnitudes were corrected for saturation following \citet{Tandon2017a}. We removed artefacts arising from saturated/bright stars and false detection at the edge, to create the final list of UVIT detected sources for each observed filter. 
We included the saturated stars in the catalogue, however their magnitudes represent the upper limit (they are brighter than these values), and their astrometry may be incorrect by a few arcseconds.
In this way, over 100 stars were detected in each image and the details of this are listed in Table~\ref{tab:cluster_obs}. 

\subsection{\textit{Gaia} data} \label{dec:gaia}

The \textit{Gaia} EDR3 data for all clusters were compiled by constraining spatial and parallax measurements. We used the r50 (radius containing half the members) mentioned in \citet{Cantat2020} to get the majority of the members with minimal contamination in the VPD. This region was used to calculate MPs in the {\sc gmm} model. We tripled the radius for running the {\sc prf} algorithm to detect more members in the outer region.
The definition and formulae of independent/derived \textit{Gaia} parameters used in this work are shown in Table~\ref{tab:Gaia parameters}.
The errors of \textsc{ra, dec, pmra, pmdec} and \textsc{parallax} are taken from \textit{Gaia} EDR3. Upper limits of the photometric errors in \textsc{g} are calculated using the `mag\_error' formula in Table~\ref{tab:Gaia parameters} (similar for errors in \textsc{gbp, grp, bp\_rp bp\_g} and {\sc g\_rp}). Errors in {\sc ruwe} and {\sc aen} are assumed to be zero.

The parallax\_cluster, ra\_cluster, dec\_cluster and radius\_cluster (as mentioned in Table~\ref{tab:cluster_details}) are used to select sources near cluster using following ADQL query:
\noindent\rule[0.5ex]{\linewidth}{0.5pt}
{\footnotesize {\fontfamily{pcr}\selectfont
\noindent 
select *
\\\textcolor{magenta}{from} gaiaedr3.gaia\_source
\\\textcolor{magenta}{where} 
\\pmra \textcolor{magenta}{is not null and} parallax \textcolor{magenta}{is not null and}
\\ABS(parallax-\textit{\textbf{cluster\_parallax}})$<$3* parallax\_error \textcolor{magenta}{and}
\\\textcolor{magenta}{contains}(\textcolor{magenta}{point}('icrs', gaiaedr3.gaia\_source.ra, gaiaedr3.gaia\_source.dec), \textcolor{magenta}{circle}('icrs', \textbf{\textit{cluster\_ra, cluster\_dec, cluster\_radius}})) = 1
}}
\noindent\rule[0.5ex]{\linewidth}{0.5pt}

\subsection{Isochrones and evolutionary tracks} \label{dec:iso_tracks}

We used {\sc parsec} isochrones \footnote{http://stev.oapd.inaf.it/cgi-bin/cmd} \citep{Bressan2012} generated for cluster metallicity and age, adopted from \citet{Dias2002} and WEBDA\footnote{https://webda.physics.muni.cz/}. We used \citet{Kroupa2001} initial mass function for the isochrones.
As the UV images would detect WDs, we included WD (hydrogen rich atmosphere, type DA) cooling curves in the CMDs. We used tracks by \citet{Fontaine2001} and \citet{Tremblay2011} for \textit{Gaia} filters \footnote{http://www.astro.umontreal.ca/~bergeron/CoolingModels/} and UVIT filters (Bergeron P., private communication). As the turnoff masses of the OCs under study range from 1.1 to 2.3 M$_{\odot}$, we included WD cooling curves of mass 0.5 to 0.7 M$_{\odot}$ \citep{Cummings2018}.
The extinction for all filters was calculated using \citet{Cardelli1989} and \citet{Odonnell1994}. We used reddened isochrones and WD cooling curves in this paper.

\section{Membership Determination} \label{sec:method}

\subsection{Gaussian Mixture Model} \label{sec:gmm}
The distribution of stars in the PM space is assumed to be an overlap of two Gaussian distributions. The sum of which can be written as,
\begin{equation}
    f(\mu|\overline{\mu_j},\Sigma_j)=\sum_{j=1}^{2} w_j \frac{exp\left[-1/2(\mu-\overline{\mu_j})^T\Sigma^{-1}_{j}(\mu-\overline{\mu_j})\right]}{2\pi\sqrt{det \Sigma_j}}
\end{equation}
\begin{equation}
    w_j>0,\qquad \sum_{j=1}^{2} w_j=1
\end{equation}
where $\mu$ is individual PM vector, $\overline{\mu_j}$ are field and cluster mean PMs, $\Sigma$ is the symmetric covariance matrix and $w_j$ are weights for the two Gaussian distributions.
The generalised formalism for the n-D case and details of fitting the Gaussian distributions to \textit{Gaia} data are available in \citet{Vasiliev2019} appendix.

We selected stars within r50 of cluster centre and removed sources with following quality filters \citep{Lindegren2018, Riello2020} to keep stars with good astrometric solutions:
\begin{equation} \label{eq:quality_GMM}
    \begin{split}
    \textsc{ruwe} > 1.4 \\
    \textsc{aen > 1.0}\quad \textsc{and} \quad  \textsc{aen\_sig > 2.0} \\
    |\textsc{parallax} - parallax\_cluster| > 3 \times \textsc{parallax\_error}
\end{split}
\end{equation}

For such sources, a {\sc gmm} is created using {\sc pmra} and {\sc pmdec}, as only these parameters have distinct Gaussian distribution for the cluster members. Two isotropic Gaussian distributions are assumed for the field and member stars, which were initialised with previously know values of cluster PM and internal velocity dispersion.
We used {\sc GaiaTools}\footnote{https://github.com/GalacticDynamics-Oxford/GaiaTools} to maximise the likelihood of the {\sc gmm} and get mean and standard deviation of the two Gaussian distributions. Simultaneously, the MPs of all stars in the field are calculated.

{\sc gmm} cannot use the other parameters provided by \textit{Gaia} EDR3 catalogue ({\sc parallax, ra, dec, g, bp\_rp}, etc.) due to their non-Gaussian distributions. {\sc gmm} does not organically account for systematic parameters leading to loss of interesting stellar systems with variability, binarity and atypical spectra. However, {\sc gmm} can convincingly give the average CMD and VPD distribution of stars in a cluster. This can be further enhanced with the inclusion of photometric and systematic information.
Hence, we used a supervised ML method to improve membership determination and utilise the non-Gaussian parameters.

\begin{figure*}
  \begin{minipage}[c]{0.72\textwidth}
    \includegraphics[width=\textwidth]{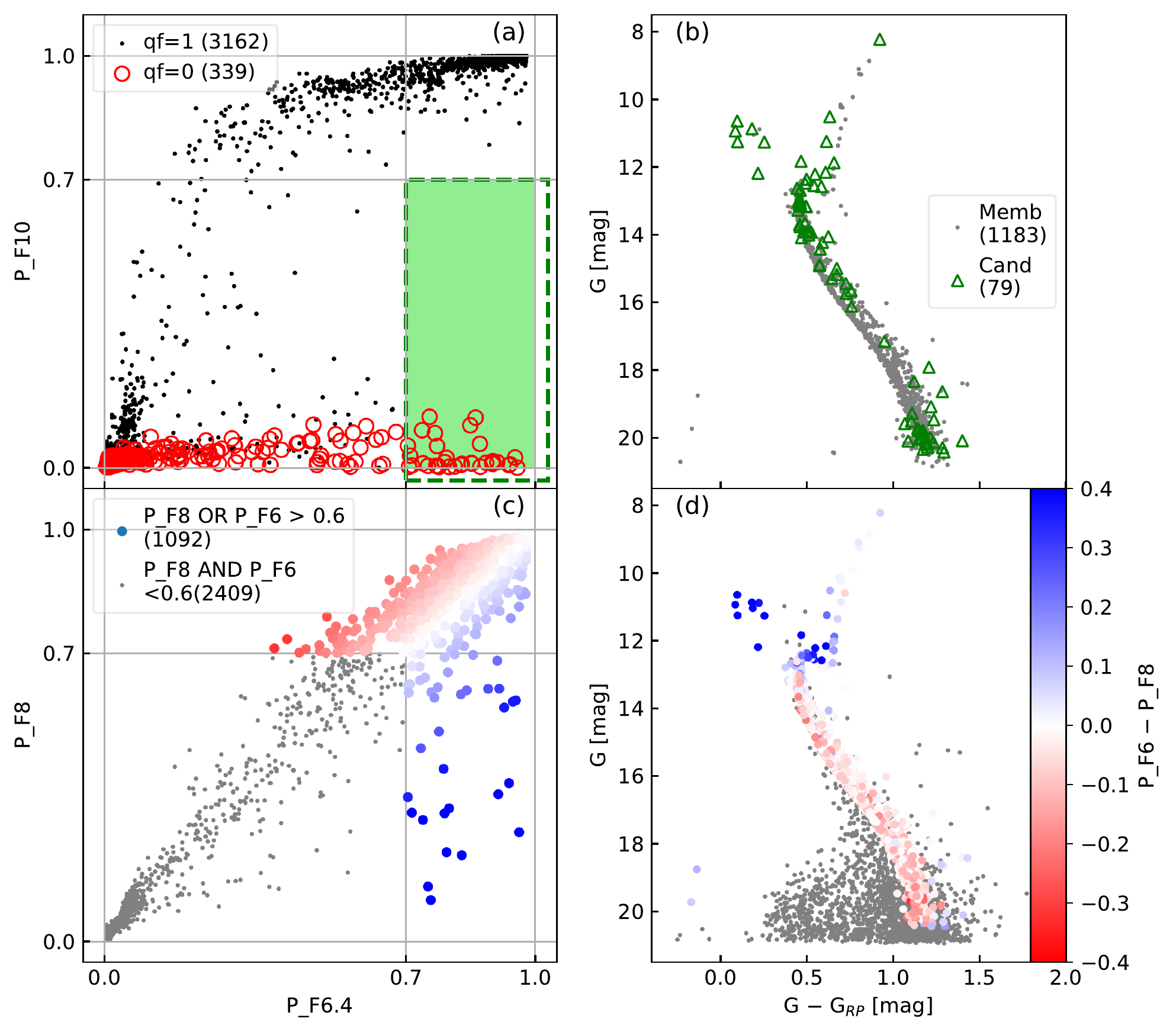}
  \end{minipage}\hfill
  \begin{minipage}[c]{0.27\textwidth}
    \caption{Comparison of different MPs from {\sc prf} feature-combinations for NGC 2682. The numbers in brackets represent the number of stars in that particular category. 
    \textbf{(a)} Comparison of F6 and F10, where black dots are good quality sources ({\sc qf} = 1) and red circles are poor quality sources ({\sc qf} = 0). The green dashed box represents sources with \(P\_F10 < 0.7 \leq P\_F6 \). 
    \textbf{(b)} CMD of NGC 2682 according to membership criteria in Eq.~\ref{eq:classif}, the grey dots are \texttt{members} (\(P\_F10 > 0.7\)), while the green triangles are the \texttt{candidates} (\(P\_F10 < 0.7 \leq P\_F6\)).
    \textbf{(c)} Comparison of F6 and F8 to demonstrate the MP dependence on CMD location. The colour is according to (P\_F6 $-$ P\_F8), as shown in the right most panel.
    \textbf{(d)} CMD of stars coloured according to (P\_F6 $-$ P\_F8). The stars with \textit{peculiar} CMD position are bluer.
    } \label{fig:P_comparison_internal}
  \end{minipage}
\end{figure*} 

\subsection{Probabilistic Random Forest} \label{sec:prf}
A random forest consists of multiple decision trees. Each decision tree consists of nodes, where the values of the features are compared with thresholds. These thresholds are optimised in the training phase by using a sample of known sources, each with a given set of features (e.g. photometry, astrometry, etc.) and labels (member, non-member). The decision trees repeatedly split the sample up until only one class of source remains at the end of each split. Finally, the average classification of all the decision trees is used, each of which is trained on different subsets of the sample.

The major drawback of traditional random forest is that there are no uncertainties (errors) assumed during these calculations. The probabilistic random forest ({\sc prf}) algorithm \footnote{https://github.com/ireis/PRF}, developed by \citet{Reis2019}, takes care of errors in the data, which is essential for any astronomical data-set.
It assumes all features and labels as probability distribution functions and out-performs traditional random forest algorithms in case of noisy data sets.

To create a training set for the {\sc prf} algorithm, we first calculated the MPs using the {\sc gmm} method. We used stars within r50 radius \citep{Cantat2020} from the cluster centre to reduce field star contamination.
The training set was created for each cluster by labelling $P\_GMM>0.5$ as members and others as non-members. The training set and testing set were created by randomly splitting the data set in 3:1 ratio, respectively.
The {\sc prf} requires the features, their errors and the known class ({\sc gmm} membership labels) as inputs for training. The output contains fractional MP for each star and the feature-importance.
After training the {\sc prf} on stars within radius r50, we applied the algorithm on the stars within 3$\times$r50 of cluster centre to increase the sample size.

We assessed the performance of the following parameters, as features which can impact the membership determination:
{\sc ra, dec, pmra, pmdec, parallax, g, g\_rp, ruwe, aen, pmR0} and many others. 
The meaning of the features used are mentioned in Table~\ref{tab:Gaia parameters}. Radial velocities (RVs) are limited to stars with {\sc g} $<15$; hence they were not used as a feature. 

We tried more than 22 feature-combinations to optimise the membership determination. We judged the different combinations by:

\begin{enumerate}[leftmargin=*]
    \item their ability to recreate MP similar to {\sc gmm} using `accuracy score' in testing phase. The accuracy score is defined as: 
        \begin{equation}    
        accuracy\  score = \frac{correctly\  predicted\  class}{total\  testing\  class}\times 100
        \end{equation}
     Although the accuracy score itself is not enough to select the final feature-combination, one can weed out poorly performing combinations.
    \item the distribution of members in VPDs (Cluster should occupy compact circular region in the VPD {e.g. Fig.~\ref{fig:CV_combined}}).
    \item the distribution of members in CMDs (Minimal contamination to the CMD, although it is a subjective judgement).  
    \item the distribution of members in PM--parallax plot (Cluster should occupy small range in both PM and distance). 
\end{enumerate}

Based on their individual merits, the notable feature-combinations are listed in Table~\ref{tab:feature_selection}.

\begin{table}
\centering
\caption{Feature-combinations used in {\sc prf} algorithm to calculate MP.}
\label{tab:feature_selection}
\begin{tabular}{L{0.7cm}L{4.7cm}C{1.5cm}}
\toprule
Name	&	Features	& 	Information	\\ \midrule
F6	& 	{\sc ra, dec, pmra, pmdec, parallax, pmR0} 	&	Astrometry	\\ \hline
F8	&	{\sc ra, dec, pmra, pmdec, parallax, pmR0, g, g\_rp} 	&	Astrometry+ Photometry	\\ \hline
F10	&	{\sc ra, dec, pmra, pmdec, parallax, pmR0, g, g\_rp, ruwe, aen} 	&	Astrometry+ Photometry+ Systematics	\\

 \bottomrule
 \end{tabular}
 \end{table} 
 
\subsection{Selection of features and membership criteria} \label{sec:feature_selection_0}

We trained {\sc prf} using 1 to 1000 trees and saw a plateau in accuracy score after 150--200 trees. As \citet{oshiro_how_2012} suggested, the optimum number of trees lies between 64--128, hence we chose 200 trees for further analysis. 
Almost all feature-combinations had an accuracy score of 92--98\%, as all were designed to select the cluster members. Hence, choosing the best combination is not trivial. 

As expected {\sc pmra, pmdec} and {\sc parallax} are important features for membership. The cluster's distribution in VPD is Gaussian, and the random forest does not completely replicate this quadratic relation between {\sc pmra} and {\sc pmdec}. Hence we created a new parameter called {\sc pmR0}, which is the separation of the source from the cluster centre in the VPD. The PM cluster centre was obtained from the {\sc gmm} results. {\sc pmR0} helped constrain the cluster distribution to a circular shape.
To test the importance of individual features, we introduced a column with random numbers as a feature. Among the \textit{Gaia} features, {\sc ra} and {\sc dec} showed very comparable feature-importance as the random column. However, upon further inspection, we found that inclusion of {\sc ra/dec} does not harm the {\sc prf} while improving the membership determination (King 2 is an example, which is the farthest cluster in our set and has smallest sky footprint). Due to known overestimation of {\sc gbp} flux in fainter and redder stars \citep{Riello2020}, we used {\sc g} and {\sc g\_rp} as features.

We added {\sc ruwe} and {\sc aen} as features to include the quality checks in {\sc prf}. This nullifies the need for manually filtering the data. The sources with large {\sc ruwe/aen} are typically binary stars, variables, extended sources or stars with atypical Spectral Energy Distribution (SED; \citealt{Lindegren2020, Riello2020, Gaia2020_summary, Fabricius2020}. As binaries and atypical SEDs are an important part of the clusters, we devised a method to keep such poor quality sources as candidates. Hereafter, we will refer to sources with {\sc qf = 1} as `good quality sources' and sources with {\sc qf = 0} as `bad quality sources'.
Feature-combination F6 uses only astrometric data (see Table~\ref{tab:feature_selection}) for the membership determination, hence it can give the MPs for poor quality sources. F10 uses astrometric, photometric and systematic parameters as features, hence it can give membership of good quality sources. Fig.~\ref{fig:P_comparison_internal} (a) shows the comparison of MPs from F6 and F10. As seen from the CMD in Fig.~\ref{fig:P_comparison_internal} (b), the bad quality sources in green region (P\_F10 $<$ 0.7 $<$ P\_F6) are very good candidate to be cluster members. For further text we define the \texttt{members, candidates} and \texttt{field}, as follows:
\begin{equation} \label{eq:classif}
    \begin{split}
    \texttt{Members} \Rightarrow {}& P\_F10 > cutoff\\
    \texttt{Candidates} \Rightarrow {}& P\_F10 < cutoff \leq P\_F6.4 \\
    \texttt{Field} \Rightarrow {}& P\_F10\  \textsc{and}\  P\_F6 < cutoff
    \end{split}
\end{equation}
\noindent In an ideal scenario without systematic errors, we would use only F6 for the membership. We recommend using the \texttt{candidate} classification for {\sc G $<$ 19} (large intrinsic errors at fainter magnitudes create spread in bottom MS). After looking at the CMDs and residual VPDs with various cutoffs, we recommend cutoff of 0.7. However, we note that the ideal cutoff varies from cluster to cluster and strongly depends on the separation of cluster--field in the VPD and ratio of field stars to cluster members.

While analysing the different feature combinations, we found that adding photometric information (F8) to astrometric information (F6) leads to lessening the MP of stars in \textit{peculiar} CMD locations. This is demonstrated in Fig.~\ref{fig:P_comparison_internal} (c) and (d). Most stars have the same MP from F8 and F6, however, the BSSs in NGC 2682 have larger P\_F6~$-$~P\_F8, due to absence of many stars in the same location in the training set. For further discussion, we will refer to (P\_F6~$-$~P\_F8) as \textit{peculiarity}. Unfortunately, other clusters do not have many BSSs, the \textit{peculiarity} can be used to distinguish between stars on the MS and sub-giant/giant branch.

\begin{figure*}
    \centering
    \includegraphics[width = 0.98\textwidth]{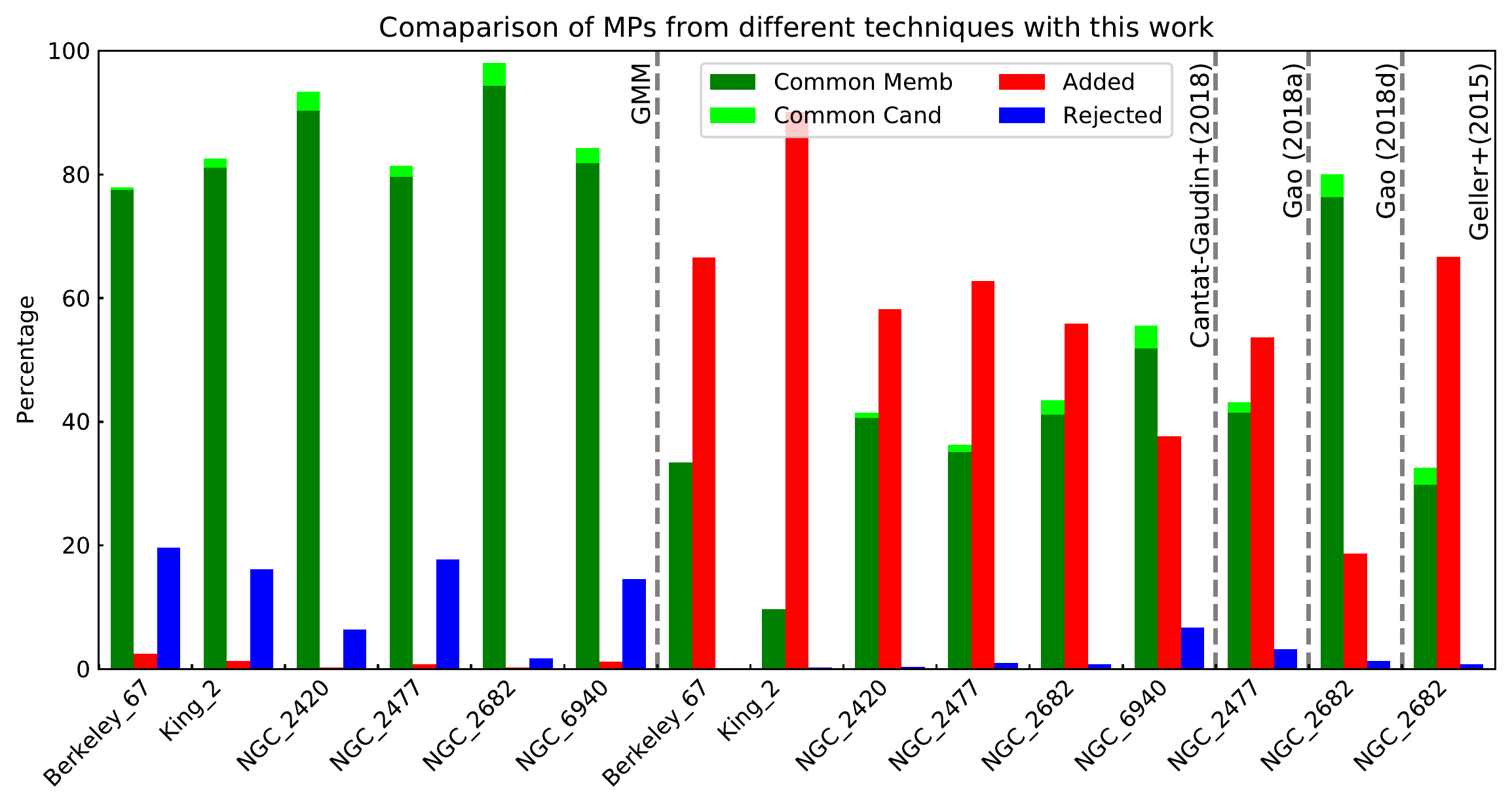}
    \caption{Grouped histogram for common \textit{candidates}, common \textit{members}, added \textit{members} and rejected stars. The totals of four classes are normalised to 100 for easy visualisation and the numbers are tabulated in Table~\ref{tab:lit_comparison}. The likely members by {\sc prf} and other techniques are represented by Common Memb (dark green) and Common Cand (lime). Red bars are stars added as \texttt{members} by {\sc prf} (classified as field by other techniques or missing from literature catalogues). Blue bars are stars rejected (classified as \texttt{field}) by {\sc prf} but these are members in other techniques. Most of the rejected stars are faint and with larger {\sc pmR0}. The dotted lines separate the comparisons with different methods and papers viz. {\sc gmm}, \citet{Cantat2018}, \citet{Gao2018b}, \citet{Gao2018c} and \citet{Geller2015}. All comparisons are done for the same field of view.}
    \label{fig:lit_comparison}
\end{figure*}

\begin{table}
    \centering
    \caption{Comparison of membership classification by {\sc prf} with {\sc gmm}, \citet{Cantat2018}, \citet{Gao2018b}, \citet{Gao2018c} and \citet{Geller2015}.}
    \begin{tabular}{lcc cc}
    \toprule
	&	Common	&	Common	&	Added	&	Rejected	\\
	&	\texttt{Memb}	&	\texttt{Cand}	&	\texttt{Memb}	&	Stars	\\
	&	[1]	&	[2]	&	[3]	&	[4]	\\ \midrule
\multicolumn{5}{c}{Comparison with {\sc gmm}}									\\
Berkeley 67	&	158	&	1	&	5	&	40	\\
King 2	&	506	&	9	&	8	&	101	\\
NGC 2420	&	354	&	12	&	1	&	25	\\
NGC 2477	&	1416	&	33	&	14	&	316	\\
NGC 2682	&	436	&	17	&	1	&	8	\\
NGC 6940	&	338	&	10	&	5	&	60	\\ \midrule
\multicolumn{5}{c}{Comparison with \citet{Cantat2018}}									\\
Berkeley 67	&	131	&	0	&	261	&	0	\\
King 2	&	104	&	0	&	968	&	3	\\
NGC 2420	&	357	&	7	&	511	&	3	\\
NGC 2477	&	1396	&	46	&	2492	&	39	\\
NGC 2682	&	502	&	28	&	681	&	9	\\
NGC 6940	&	399	&	29	&	290	&	52	\\ \midrule
\multicolumn{5}{c}{Comparison with \citet{Gao2018b}}									\\
NGC 2477	&	1695	&	67	&	2193	&	133	\\ \midrule
\multicolumn{5}{c}{Comparison with \citet{Gao2018c}}									\\
NGC 2682	&	950	&	46	&	233	&	16	\\ \midrule
\multicolumn{5}{c}{Comparison with \citet{Geller2015}}									\\
NGC 2682	&	365	&	34	&	817	&	10	\\ \bottomrule
\multicolumn{5}{c}{\footnotesize [1] Classified as \texttt{members} by both {\sc prf} and other techniques}									\\
\multicolumn{5}{c}{\footnotesize [2] Members of other techniques classified as \texttt{Candidates} by {\sc prf}}									\\
\multicolumn{5}{c}{\footnotesize [3] Added \texttt{members} by {\sc prf}, which are not members in other catalogues}									\\
\multicolumn{5}{c}{\footnotesize [4] Members from other techniques classified as \texttt{field} by {\sc prf}}									\\
\end{tabular}
    \label{tab:lit_comparison}
\end{table} 

\subsection{Comparison with literature}

Fig.~\ref{fig:lit_comparison} shows the comparison of {\sc prf} with {\sc gmm}, \citet{Cantat2018}, \citet{Gao2018b} and \citet{Gao2018c}. The actual numbers of different types of stars are listed in Table~\ref{tab:lit_comparison}. Although intuitive, the meaning of `added', `rejected' etc. is given in Table~\ref{tab:lit_comparison} footnote. Although \citet{Lindegren2020} warns against direct crossmatch between DR2 and EDR3 due to changes in epochs, the astrometric shift for cluster members is $<$ 10 mas. All comparisons were done over the same field of view.

{\sc gmm} and {\sc prf} used different set of parameters. As seen Fig.~\ref{fig:lit_comparison}, the classification by {\sc prf} was similar to {\sc gmm}. The accuracy score (reproducibility) of {\sc prf} was between 90--99\% for the six clusters. However, there are some differences, which are expected and embraced. The major difference was seen in the rejection of {\sc gmm} members (2--25\%), most of which were in {\sc G} $>$ 19 mag region. Among stars brighter than 19 mag, the percentage of rejected stars drops to 0--4\%, almost all having poor astrometric solutions ({\sc ruwe $>1.2$}). 

\citet{Cantat2018} used clustering in the ($\mu_{\alpha},\mu_{\delta}, \pi$) space to identify the members using \textit{Gaia} DR2 data. They selected the stars with parallax within 0.3 mas and PM within 2 mas year$^{-1}$ of the cluster mean. The probabilities were calculated using {\sc upmask}, an unsupervised clustering algorithm. {\sc prf} has identified significantly more (260-2500) new members compared to \citeauthor{Cantat2018}. All the added stars have acceptable CMDs, VPDs and PM--parallax distributions. As the magnitude limit of \citeauthor{Cantat2018} catalogue was 18, many new members are added in the fainter end of the MS. The rejected members (0--52) typically have either high {ruwe}/low {\sc pmR0} or low {\sc ruwe}/high {\sc pmR0}. As the added stars far outnumber the rejected stars, this is a good optimisation.

\citet{Gao2018b} used \textit{Gaia} DR2 to determine membership of NGC 2477 (and three other clusters) using a {\sc gmm}. Although there are 1695 common members, {\sc prf} has added 2193 stars and rejected 133 stars. Majority of added stars are near the cluster parallax and {\sc pmR0} $<$ 1.5 mas yr$^{-1}$. The rejected stars are again typically results of {\sc ruwe/aen} and {\sc pmR0} trade-off.

\citet{Gao2018c} utilised a random forest of 11 \textit{Gaia} parameters ({\sc ra, dec, parallax, pmra, pmdec, g, gbp, grp, bp\_rp, bp\_g and g\_rp }) to calculate the MPs of NGC 2682. \citeauthor{Gao2018c} did not remove stars with high systematic errors, and the random forest algorithm did not incorporate uncertainties in the astrometric data. The use of EDR3 data, inclusion of errors and using the F6 and F8 feature-combinations has led to 233 more \texttt{members}. 
\citet{Geller2015} calculated the MPs in NGC 2682 using a combination of RV measurements (up to 40 yr baseline) and previous PM data \citep{Yadav2008}. Their catalogue is magnitude limited due to its spectroscopic nature. Among the crossmatched \citeauthor{Geller2015} members, we classify 3\% stars as \texttt{field}, due to larger {\sc ruwe} or different PM/{\sc parallax}.

Comparison with previous literature confirms that F10 membership is adequate for membership, and we will use F10 as primary membership criteria.
Due to the limitations (systematic and statistical errors) in \textit{Gaia} EDR3, we included the \texttt{candidate} classification to account for the likely cluster members (using F6). \footnote{literature comparison for all above clusters and methods is available as Fig.~\ref{fig:rVC_be67} to Fig.~\ref{fig:gao_comp} in only the arXiv version.}
We used both \texttt{members} (selected from F10)  and \texttt{candidates} (selected from F10 and F6) for further analysis. 

Importantly, our method was able to add a significantly large number of stars in all the clusters, ranging from 200--2500 stars per cluster (Table~\ref{tab:lit_comparison}). Therefore, this is a significant improvement over the previous studies using \textit{Gaia} DR2, mostly in the faint MS. This will certainly help in the detailed analysis of the clusters and locate interesting candidates that are bright in the UV.

\section{Results}\label{sec:results}

\begin{table*}
\centering
\caption{The cluster parameters as derived from the \texttt{members} (see Eq.~\ref{eq:classif}) of the six clusters.
} 
\label{tab:cluster_parameters}
\begin{tabular}{lcc ccc cc} 
\toprule
Cluster			&	Berkeley 67	&	King 2	&	NGC 2420	&	NGC 2477	&	NGC 2682	&	NGC 6940	\\ \midrule
Total stars in 3$\times$r50			&	4962	&	4343	&	1604	&	37649	&	3501	&	99769	\\
\texttt{Members}			&	392	&	1072	&	868	&	3888	&	1183	&	689	\\
\texttt{Candidates}			&	33	&	46	&	47	&	174	&	79	&	43	\\ \midrule
ra\_mean		[degree] $^{\alpha}$	&	69.471	&	12.727	&	114.603	&	118.048	&	132.844	&	308.632	\\
dec\_mean		[degree] $^{\alpha}$	&	50.743	&	58.186	&	21.577	&	$-$38.534	&	11.827	&	28.300	\\
pmra\_mean		[mas yr$^{-1}$] $^{\alpha}$	&	2.28$\pm$0.28	&	$-$1.43$\pm$0.27	&	$-$1.22$\pm$0.30	&	$-$2.43$\pm$0.26	&	$-$10.96$\pm$0.33	&	$-$1.96$\pm$0.15	\\
pmdec\_mean		[mas yr$^{-1}$] $^{\alpha}$	&	$-$1.42$\pm$0.22	&	$-$0.85$\pm$0.35	&	$-$2.05$\pm$0.26	&	0.90$\pm$0.27	&	$-$2.91$\pm$0.29	&	$-$9.44$\pm$0.16	\\
Stars with RV			&	2	&	1	&	6	&	16	&	39	&	15	\\
RV\_mean		[km s$^{-1}$] $^{\alpha}$	&	$\sim$ $-$1	&	$\sim$ $-$41	&	73$\pm$2	&	8$\pm$4	&	34$\pm$4	&	8$\pm$4	\\ \midrule
parallax\_mean		[mas] $^{\alpha}$	&	0.43$\pm$0.29	&	0.15$\pm$0.39	&	0.38$\pm$0.26	&	0.64$\pm$0.33	&	1.16$\pm$0.28	&	0.95$\pm$0.19	\\
distance from isochrone		[pc]	&	2023	&	5749	&	2512	&	1514	&	848	&	1000	\\
R\_core		[\arcmin] $^{\beta}$	&	1.3	&	0.5	&	1.2	&	6.4	&	6.3	&	2.2	\\
R\_core 		[pc]	&	0.76	&	0.84	&	0.88	&	2.82	&	1.55	&	0.64	\\
\toprule															
\multicolumn{7}{c}{ $^{\alpha}$ The means and errors are mean and standard deviations of Gaussian fit to the \texttt{member} parameters.}															\\
\multicolumn{7}{c}{ $^{\beta}$ Projected R\_core is calculated using distance from isochrone fits}			
\end{tabular}
\end{table*} 

\subsection{The Catalogues} \label{sec:catalogue}
The results of this study are presented in the form of seven catalogues, a membership catalogue and six catalogue of UVIT photometry. 
The membership catalogue (for sources with $P\_F6\ \textsc{or}\ P\_F10 > 0.1$) contains \textit{Gaia} EDR3 astrometry and photometry (\textsc{ra, dec, g, g\_rp}), MPs (P\_GMM, P\_F6, P\_F8 and P\_F10), quality filter (\textsc{qf}) and membership classification (M: \texttt{member}, C: \texttt{candidate} and F: \texttt{Field}). The example of the catalogue is given in Table~\ref{tab:cat_Gaia} (full version is available online).
Table~\ref{tab:cat_UV} shows the example of UV catalogue of NGC 6940, which is observed in F169M filter. The full catalogues of six clusters are available online. 
The catalogues contain R.A.(J2016), Dec.(J2016), UV magnitudes, magnitude errors, MPs (P\_F10 and P\_F6) and membership classification (M: \texttt{member}, C: \texttt{candidate} and F: \texttt{Field}). We have included saturated stars in the catalogue, whose magnitudes give the upper bound to the actual numerical value.

We cross-matched the \textit{Gaia} and UVIT catalogues with a radius of 1$^{\prime\prime}$, to get merged catalogues using \textsc{topcat}\footnote{http://www.star.bris.ac.uk/~mbt/topcat/}. We checked for crowding and issues during cross-matching process (e.g. duplicity), but both \textit{Gaia} and UVIT catalogue showed an insignificant number of stars within 1$^{\prime\prime}$ of each other (2 for all UVIT detections and $<$ 0.4\% in \textit{Gaia} detections). These merged catalogues were used for further analysis.
 
\subsection{Cluster properties} \label{sec:cluster_properties}
We derived following mean cluster properties by fitting Gaussian distribution to the \texttt{members}: R.A., Dec, parallax, PM and RV. Additionally, we included distances calculated by isochrone fitting.
We removed a few outliers while calculating the mean parallax and RV.
We fitted King's surface density profile to cluster surface density. 
\begin{equation}
    \rho(R) = F_{bg}+ \frac{F_{0}}{1+(R/R_{core})^2}
\end{equation}
\noindent where $F_{bg}$ is background counts, $F_{0}$ is count in the bin, R is the RMS of each bin limits (in degree) and $R_{core}$ is the core radius. 
We binned the \texttt{members} such that the bin area was constant for each bin in the spatial plane. This method decreased the thickness of bin as we moved outwards from the cluster centre. The smallest bin width was kept equal to the mean separation between nearby \texttt{members}, and the rest of the bins were scaled accordingly. The $F_{bg}$ was assumed to be nil for the profile fitting. 

All the parameters are tabulated in Table \ref{tab:cluster_parameters}. \textit{Gaia} EDR3 sources near each cluster are are divided into three subsets: \texttt{members, candidates} and \texttt{field}. The VPDs and CMDs of all clusters for these individual subsets are shown in Fig.~\ref{fig:CV_combined}.
The spatial distribution, VPD, CMD, {\sc g} vs MP etc. for clusters is shown in Fig.~\ref{fig:rVC_be67} to Fig.~\ref{fig:rVC_6940}. 
For each cluster, Fig. (a) shows the spatial distributions of \texttt{members} and non-\texttt{member} population. 
Fig. (b) and (c) show the distribution of GMM and F10.3 probabilities as a function of \textsc{g}. We expected clear separation between \texttt{members} for bright stars, which is seen in all the clusters up to 16--18 mags. 
Fig. (d) shows the distribution parallax as a function of \textsc{g}. All clusters, except King 2, show a peak in parallax for the \texttt{member} stars. 
Fig. (e) shows the King's surface density profile fitted to \texttt{members}' surface density.
Fig. (f) shows the histogram of F10 MP.
Fig. (g) and (k) show the VPD and CMD for all stars.
Fig. (h) and (l) show the VPD and CMD for \texttt{members}.
Fig. (i) and (m) show the VPD and CMD for \texttt{candidates}. 
Fig. (j) and (n) show the VPD and CMD for \texttt{field} stars.

\begin{figure*}
    \centering
    \begin{tabular}{c}
    \includegraphics[width=0.97\textwidth]{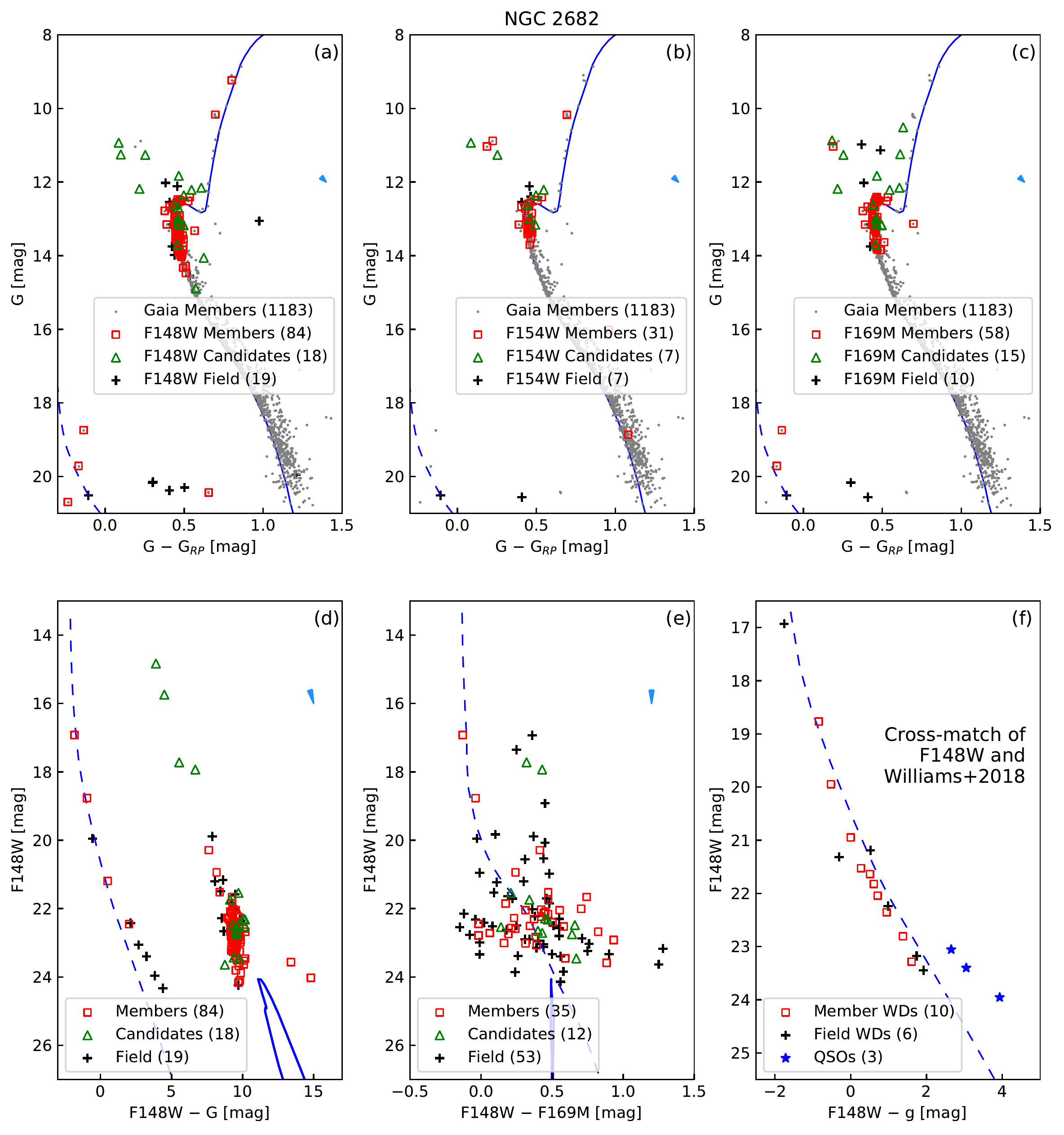}
     \end{tabular}
     \caption{The CMDs of NGC 2682 with UVIT and \textit{Gaia} photometry. Gray dots are \textit{Gaia} EDR3 \texttt{members} according to Eq.~\ref{eq:classif}. Red squares, green triangles and black crosses are \texttt{members}, \texttt{candidates} and \texttt{field} stars detected in particular filters.
     The blue line is an isochrone with $Log(Age)=9.6, Distance Modulus=9.64, E(B-V)=0.05\ mag, [M/H]=0\ dex$; the dashed blue line is WD cooling curve for a 0.5 M$_{\odot}$ WD; the light blue arrows represent the reddening vectors with direction and magnitude in each CMD.
    \textbf{(a), (b) and (c)} The optical CMDs with stars detected in F148W, F154W and F169M filters respectively.
    \textbf{(d)} The UV--optical CMDs with F148W and {\sc g} filters. 
    \textbf{(e)} The UV CMD with F148W $-$ F169M colour.
    \textbf{(f)} UV--optical CMD of sources cross-matched with \citet{Williams2018} catalogue of possible WDs. The CMD also shows three quasars (blue stars) detected by F148W filter.
    }
     \label{fig:CMD_2682}
\end{figure*} 

\begin{figure*}
    \centering
    \begin{tabular}{c}
    \includegraphics[width=0.97\textwidth]{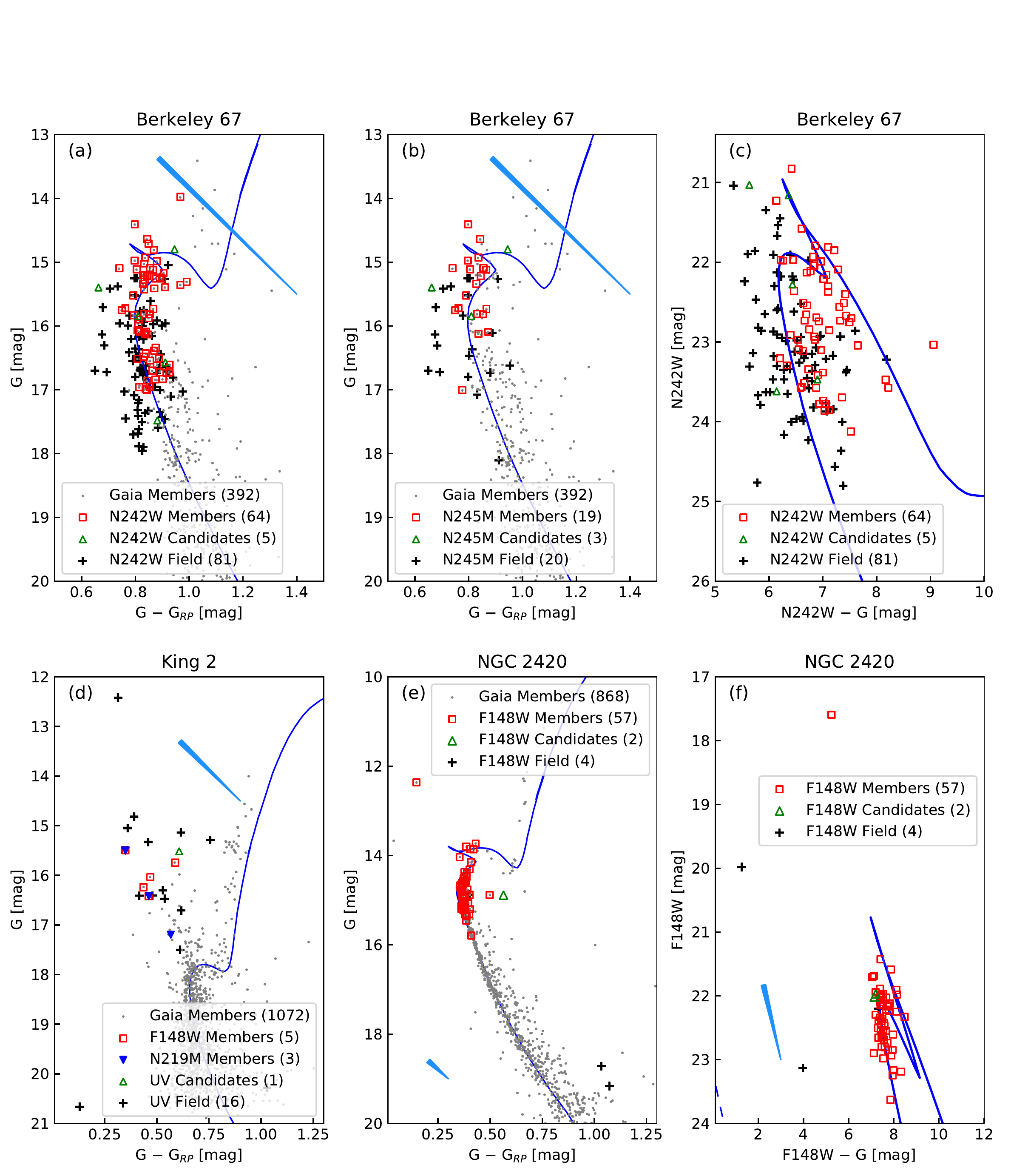} \\
    \end{tabular}
    \caption{\textbf{(a), (b) and (c)} The optical and optical--UV CMDs of Berkeley 67.
    The markers are same as Fig.~\ref{fig:CMD_2682}. The isochrones for Berkeley 67 has $Log(Age)=9.2, Distance Modulus=11.53, E(B-V)=0.8\ mag, [M/H]=0\ dex$.
    \textbf{(d)} Optical CMD of King 2. with F148W detected \texttt{members} denoted by red squares and N219M detected \texttt{members} denoted by blue filled triangles. 
    The isochrones for King 2 has $Log(Age)=9.7, Distance Modulus=13.8, E(B-V)=0.45\ mag, [M/H]=-0.4\ dex$.
    \textbf{(e) and (f)} The optical and UV--optical CMDs of NGC 2420. 
    The markers are same as Fig.~\ref{fig:CMD_2682}. The isochrones for NGC 2420 has $Log(Age)=9.3, Distance Modulus=12.0, E(B-V)=0.15\ mag, [M/H]=-0.4\ dex$; WD cooling curve is for 0.6 M$_{\odot}$ WD.}
    \label{fig:CMD_Be67}
\end{figure*} 

\begin{figure*}
    \centering
    \begin{tabular}{c}
    \includegraphics[width=0.97\textwidth]{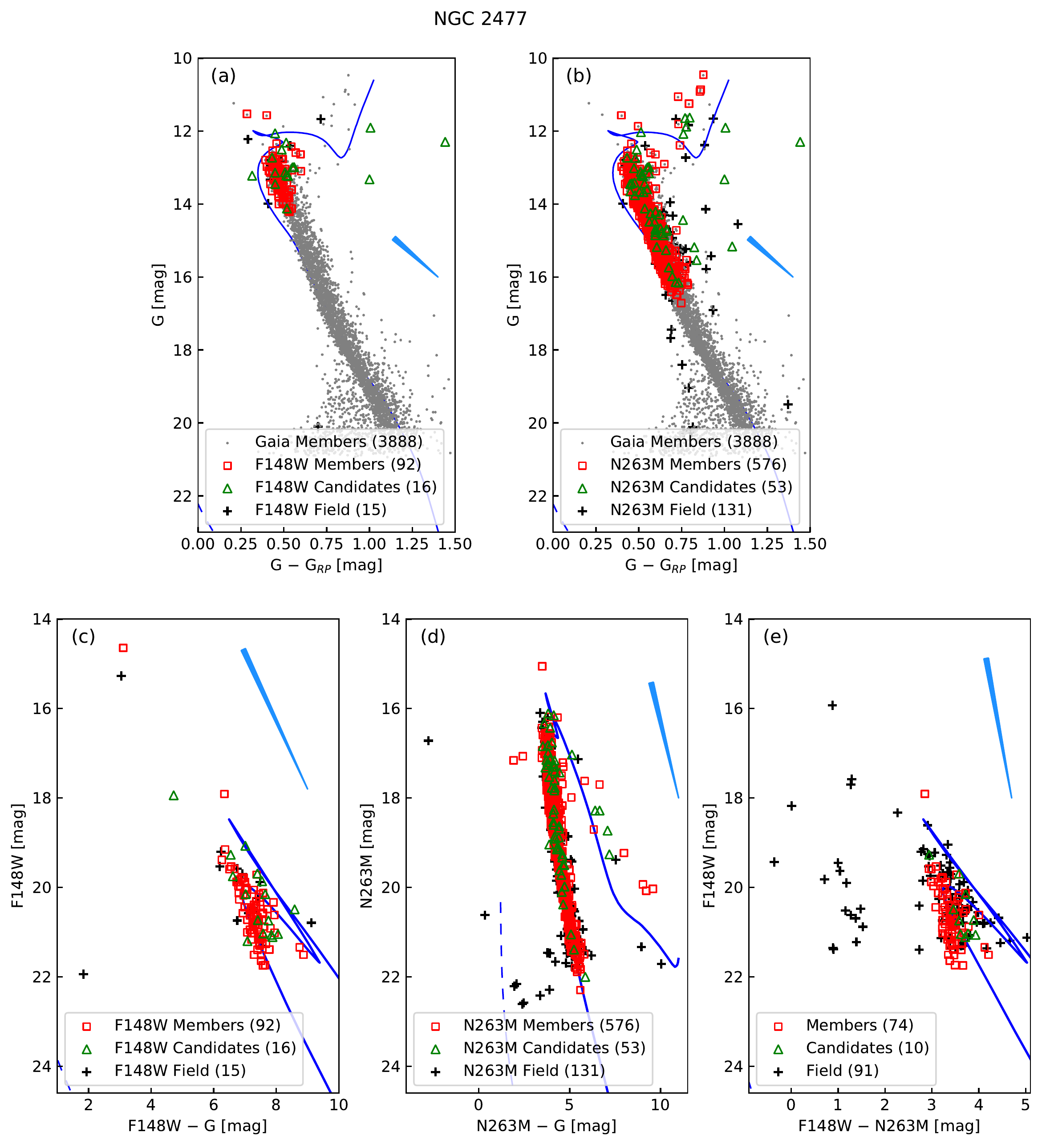} \\
    \end{tabular}
    \caption{\textbf{(a) and (b)} The optical CMDs of NGC 2477. 
    \textbf{(c), (d) and (e)} The UV--optical and UV CMDs of NGC 2477. 
    The markers are same as Fig.~\ref{fig:CMD_2682}. The isochrones is for $Log(Age)=8.9, Distance Modulus=10.9, E(B-V)=0.4\ mag, [M/H]=0\ dex$; WD cooling curve is for a 0.7 M$_{\odot}$ WD.}
    \label{fig:CMD_2477}
\end{figure*} 

\begin{figure*}
    \centering
    \begin{tabular}{c}
    \includegraphics[width=0.97\textwidth]{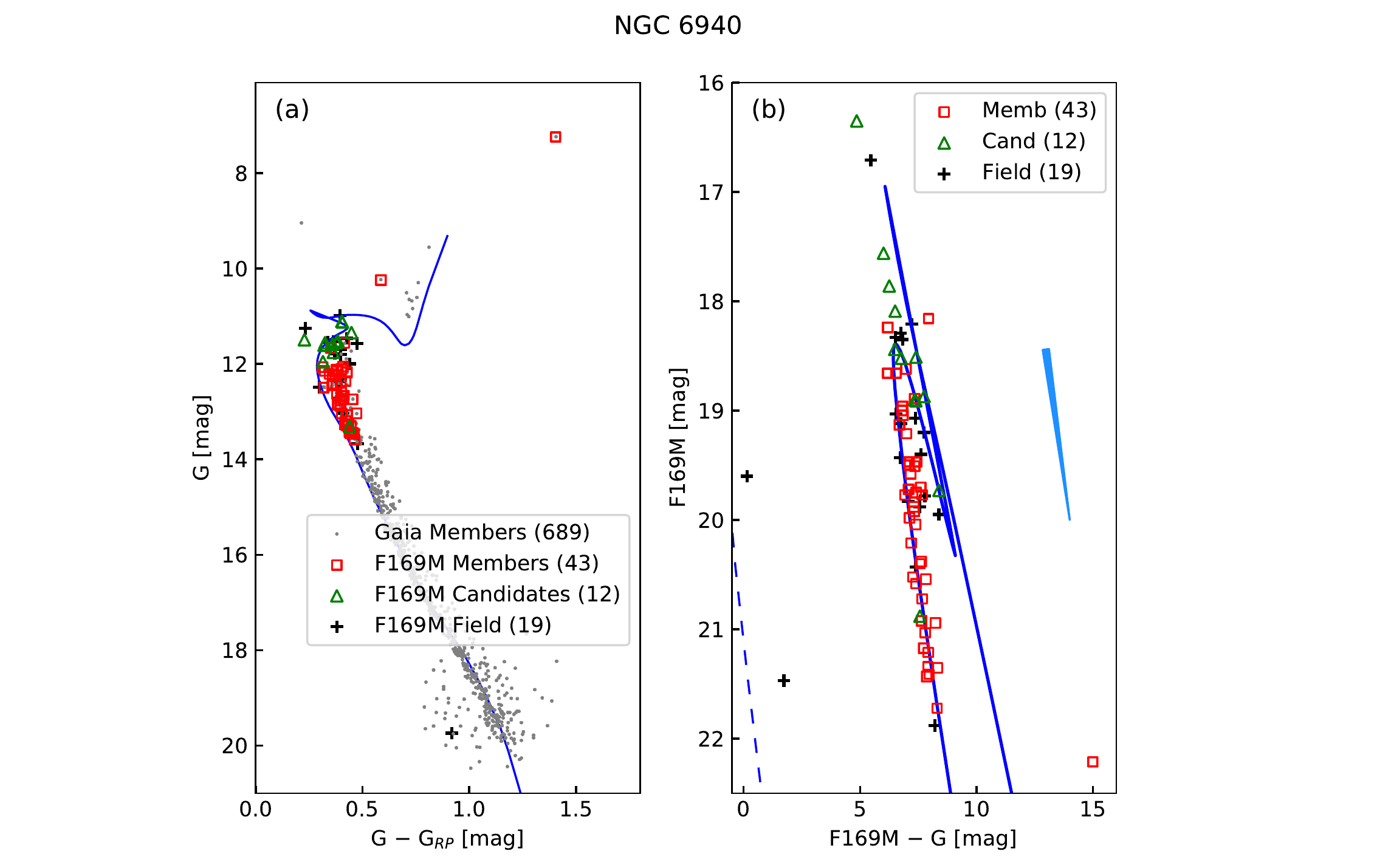} \\
    \end{tabular}
    \caption{\textbf{(a) and (b)} The optical and UV--optical CMDs of NGC 6940. 
     The markers are the same as Fig.~\ref{fig:CMD_2682}. The isochrones is for $Log(Age)=9.0, Distance Modulus=10.0, E(B-V)=0.2\  mag, [M/H]=0.0\ dex$; WD cooling curve is for a 0.6 M$_{\odot}$ WD.}
    \label{fig:CMD_6940}
\end{figure*} 


\subsection{Individual clusters} \label{sec:2682}

\textbf{NGC 2682} has a prominent binary sequence, red giant (RG) branch and BSS population (Fig.~\ref{fig:CV_combined}). We detected many \texttt{candidates}  as BSSs, MS stars and a few RG stars. The MS \texttt{candidates} typically have $\textsc{qf}=0$. 
The optical CMDs of stars detected in the FUV filters are shown in Fig.~\ref{fig:CMD_2682} (a), (b) and (c). The CMDs contain all \textit{Gaia} \texttt{members}, UVIT detected sources, isochrone and WD cooling curve of 0.5 M$_{\odot}$.
Overall, we detected 84, 31 and 58 \texttt{members} in F148W, F154W and F169M respectively. 
Fig.~\ref{fig:CMD_2682} (d) shows the UV--optical CMD of sources cross-matched between \textit{Gaia} EDR3 and F148W. The turnoff of the isochrone lies at 24 mag, which is the limiting magnitude of F148W observations. 
All the stars on optical MS are located above the turnoff in UV--optical CMD.
Hence, as previously seen in \citet{Jadhav2019}, almost all MS stars detected in optical CMD have FUV excess. 
The photometry presented here is two magnitudes fainter than \citet{Jadhav2019} in F148W, and we have detected a good number of MS stars in the 22 to 24 magnitude range, with FUV excess.
Fig.~\ref{fig:CMD_2682} (e) shows the F148W, (F148W $-$ F169M) CMD (the MS/turnoff is the vertical line at $\sim$ 0.5 colour).
The sources show a large spread in FUV colour, ranging from 0.0 to 1.3 mag.

FUV images can be used to detect WDs. However, as they are faint in the optical wavelengths, \textit{Gaia} is not suitable to detect them. In order to effectively identify the WDs, we cross-matched the F148W detected sources with the WD catalogue of \citet{Williams2018}. Fig.~\ref{fig:CMD_2682} (f) shows the CMD of all cross-matched sources and the WD cooling curve. The membership information and g-band photometry (for this subplot alone) are taken from \citet{Williams2018}. F148W has detected ten member WDs, six field WDs and three quasars. All these sources follow the WD cooling curve.

\textbf{Berkeley 67}'s VPD (Fig.~\ref{fig:CV_combined}) shows that the mean cluster motion and mean field motion are within a few mas year$^{-1}$ of each other.
The cluster has only $\sim$ 400 \texttt{members}; hence there is not much over-density in the VPD. Thus, it is particularly challenging to determine the membership for Berkeley 67. The CMD shows a large spread in the MS. We suspect the large spread is the result of the differential reddening in the cluster region, whose effect is enhanced by the high extinction towards the cluster. 
UVIT images of Berkeley 67 in F148W (exp. time = 2683 s) and F169M (exp. time = 1317s) filters detected no \texttt{member} stars. Therefore, the cluster does not have any FUV bright members.
N242W and N245M images detected 64 and 19 \texttt{members} respectively. We did not detect any BSS.
Fig.~\ref{fig:CMD_Be67} (a) and (b) show the \textit{Gaia} CMDs of these NUV detected \texttt{members}. We observe turnoff stars in both NUV filters, with the wider N242W filter going till 17 mag.
N242W filter detects two RGs which are rarely detected in the UV regime.
Fig.~\ref{fig:CMD_Be67} (c) shows the N242W, (N242W $-$ {\sc g}) CMD for Berkeley 67, which again confirms the large scatter.

\textbf{King 2} is the farthest cluster included in this work. This is evident from the small apparent core radius (0.$^\prime$5) and high feature-importance for {\sc ra} and {\sc dec}. 
The parallax measurements are unreliable at the distance of $\sim$5 kpc. The cluster and field centres in the VPD are very close and  hence, there may be contamination from field stars, among the members. 
Nevertheless, there are many BSSs and red clump stars present in the cluster (Fig.~\ref{fig:CV_combined}).
We detected 5 and 3 \texttt{members} each inF148W and N219M filter respectively (Fig.~\ref{fig:CMD_Be67} (d)). This 5 Gyr old cluster located at a large distance has the MS turnoff at 18 mag (in {\sc g}-band) and hence we detected only the brightest of BSSs in UV. Overall, there are 5 \texttt{member} BSSs and 1 \texttt{candidate} BSS in UV images. Two of the BSSs have both FUV and NUV detections. We detect one blue and faint ({\sc g}-band) object, which is likely a foreground WDs.

\begin{figure*} 
\centering
\begin{tabular}{c}
\includegraphics[width=0.95\textwidth]{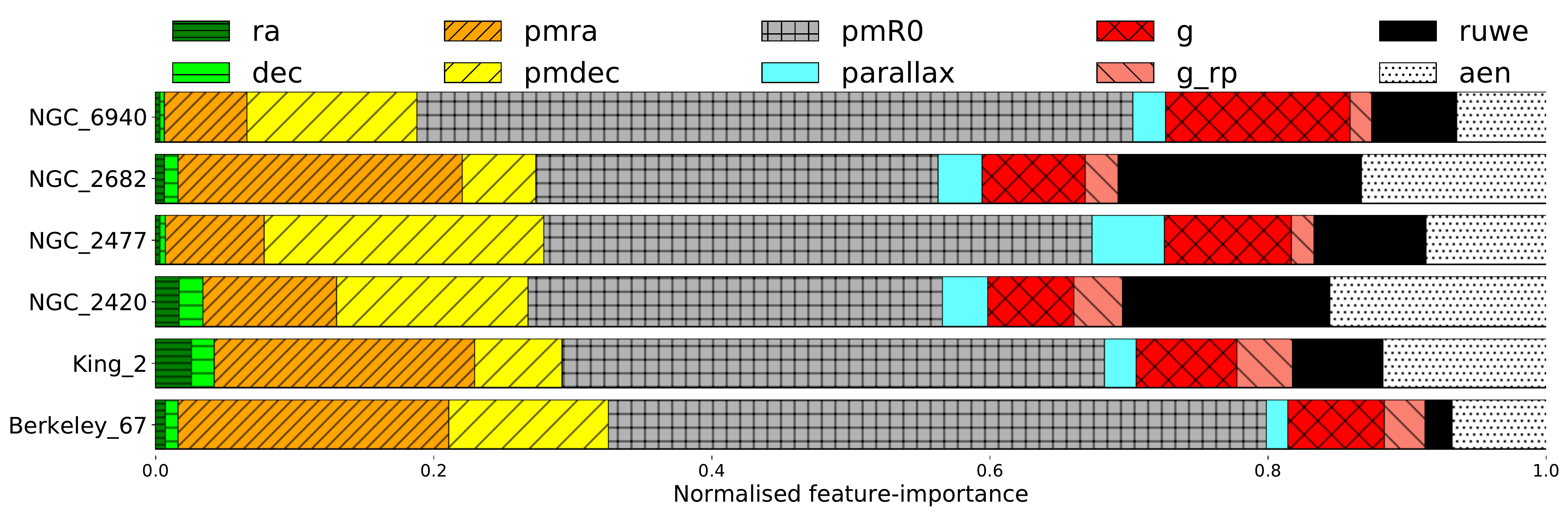}
    \end{tabular}
    \caption{Normalised feature-importance for F10 for all six OCs.}
    \label{fig:feature_importance}
\end{figure*}

\textbf{NGC 2420} lies in a relatively less dense region with the field to member stars ratio of $\sim$ 1 (Fig.~\ref{fig:CV_combined}). It has a clearly defined binary sequence and a RG branch, with a few BSSs.
The UVIT image has stars up to two magnitudes below the MS turnoff, including a BSS. The UV CMD shows that the \texttt{candidates} are located close to the turnoff, though they are much fainter in the optical CMD, suggesting a brightening in the UV. However, the excess UV flux is not as prominent as NGC 2682. Similarly, a few \texttt{members} also show UV excess flux, but the change in the magnitude is not as large as in NGC 2682.

\textbf{NGC 2477} is a very dense cluster in a high stellar density region (38000 stars and 3900 \texttt{members} in 18 {\arcmin} radius). The \textit{Gaia} CMD shows a well defined binary sequence, RC stars and $\sim$ 5 BSSs. The turnoff has a large spread in {\sc g}-band. There is spread in the \texttt{members} below 18 mag, indicating that the probability cutoff should be slightly higher than 0.7.
We detected 92 and 576 \texttt{members} in F148W and F263M respectively (Fig.~\ref{fig:CMD_2477} (a) and (b)). Fig.~\ref{fig:CMD_2477} (c) shows the F148W, (F148W $-$ {\sc g}) CMD with 2 BSSs and MSTO stars.
Fig.~\ref{fig:CMD_2477} (d) shows the N263M, (N263 $-$ {\sc g}) CMD with a large range in NUV magnitude consisting of red clump stars and MS stars.
Fig.~\ref{fig:CMD_2477} (e) shows the F148W, (F148W $-$ N263M) CMD, here we see the turnoff stars and many \texttt{field} stars with bluer UV colour. 

\textbf{NGC 6940} is situated in a very dense stellar environment (stellar density is 13 times that of NGC 2682 neighbourhood). The cluster is well separated from the central field in the VPD and has a significant parallax ($\sim$ 1 mas), hence it is easy to extract (Fig.~\ref{fig:CV_combined}). The CMD shows a clear binary sequence and red clump stars. NGC 6940 also has a relatively broad MS with a spread near its turnoff, although less prominent compared to NGC 2477.
We detected \texttt{members} up to 2 magnitudes below turnoff (Fig.~\ref{fig:CMD_6940} (a)) including a giant. We did not detect any BSS in this 1 Gyr old cluster.


\section{Discussion} \label{sec:discussion}
\subsection{Membership Determination} \label{sec:membership_discussion}

There are multiple ways of determining memberships. 
The choice of method is mainly dictated by the aim of the study. 
Any simple method such as VPDs for membership estimation is adequate for studies requiring the estimation of cluster parameters such as mean PM, age and distance. 
Here, our objective is to identify UV bright member stars in OCs, that could be in non-standard evolutionary stages. This requires the implementation of rigorous methods to assign membership to such stars, as discussed below:

\textbf{Feature-importance:} 
The {\sc prf} method gives the importance of each feature as one of the results during the training phase. The normalised feature-importance is shown in  Fig.~\ref{fig:feature_importance}.
The most important features are {\sc pmR0, ruwe, aen, pmra, pmdec} and {\sc g}.
The distance of the stars from the cluster centre in the VPD, {\sc pmR0}, is an important feature as expected.
The high importance of {\sc ruwe/aen} is because they are cutoffs established during the training of the algorithm.
The {\sc g} importance is aided by the dependence of all errors on the magnitude. 
The {\sc ra/dec} importance is more for King 2 and NGC 2420 when compared to other clusters. They have core radii of 0.$^\prime$5--1.$^\prime$2, while other clusters are typically larger. The smaller spatial distribution of members is causing an increase in the importance of {\sc ra/dec}. As expected, the importance of {\sc parallax} increases for nearby clusters (NGC 2420, NGC 2477, NGC 2682 and NGC 6940). 

\textbf{Efficacy in various environments:}
The {\sc prf} technique works for OCs with a diverse cluster-members to field-stars ratio (0.01 to 1.3), thereby helping in efficient detection of members. 
The presence of systematic errors and including CMD locations through magnitude and colour tends to remove poor quality as well as \textit{peculiar} stars. 
Therefore, we introduced the \texttt{candidate} classification to list such stars.
For these six clusters, we found the \texttt{candidates} to \texttt{members} ratio to be 0.04--0.08. 

\textbf{Versatility of the technique:}
The algorithm is adaptable, and one can choose a particular feature-combination depending on the requirements. For example, for a statistical study of clusters, a feature combination with \textsc{ra, dec} and \textsc{parallax} would be enough. To find \textit{peculiar} stars in the CMD, one could measure the difference between F6 ({\sc ra/dec, pmra/pmdec, parallax} and {\sc pmR0}) and F8 (F6 + {\sc g} and {\sc g\_rp}).
\textit{Peculiar} stars typically have lower P\_F8.
The {\sc prf} technique can also be applied to any data-set besides \textit{Gaia} EDR3. Moreover, the inclusion of {\sc ruwe/aen} as features indicates any systematic terms, if present in any other data-set, can also be incorporated in the algorithm.

\textbf{Classification of BSSs:}
In the field of NGC 2682, there were 10 potential BSSs (bluer and brighter than the turnoff). {\sc prf} classified two as \texttt{members}, six as \texttt{candidates} and two as \texttt{field}. Many of these stars are photometric variables or binaries \citep{Geller2015}, which can lead to high $\textsc{ruwe}$ and hence classification as \texttt{candidates}. The two field stars have cluster parallax and RV (not considered as a membership criteria in {\sc prf}). However, they have larger {\sc pmR0} leading to their rejection as members. For such stars with large PM deviation from the cluster mean, deeper RV measurements and accurate parallax will be useful in constraining membership.
High \textit{peculiarity} in combination with high {\sc ruwe} of the BSSs is the reason for these stars to be categorised as \texttt{candidates}. 
Hence, the technique (Eq~\ref{eq:classif}) is capable of selecting BSSs (albeit as \texttt{candidates}).

\textbf{Existence of \texttt{Candidate} class:}
All the cluster \texttt{candidates} lie near the cluster centre in the VPDs. Their number increases as they get fainter; this mirrors the fact that the systematic errors in \textit{Gaia} EDR3 are larger for fainter stars. The CMD of NGC 6940 \texttt{candidates} (Fig.~\ref{fig:CV_combined}) shows that the majority of them lie on the binary sequence. A similar but lesser effect is seen in NGC 2477 and NGC 2682. Binary systems are known to produce high {\sc ruwe} values due to variability or unsymmetrical PSF \citep{Deacon2020}, hence they can have low P\_F10 and get classified as \texttt{candidates}. 

\textbf{Detection of \textit{peculiar} stars using multiple feature-combinations:}
In Fig.~\ref{fig:P_comparison_internal} (e), we compared the MPs with and without \textsc{g} and \textsc{g\_rp} as features (F8 and F6 respectively).
F6 has no knowledge of CMD positions, so it uses only spatial location and velocity to classify stars. However, F8 selects stars with common CMD positions and rejects stars with uncommon CMD positions.
This effect is demonstrated by the positive values of $P\_F6-P\_F8$ for BSSs in NGC 2682 (Fig.~\ref{fig:P_comparison_internal} (d)). 
We refer to large $P\_F6-P\_F8$ as \textit{peculiarity}.
Such \textit{peculiarity} can be seen for the BSSs in NGC 2682, NGC 2477 and King 2. However, for King 2 majority of stars bluer than {\sc bp\_rp} $<$ 1.1 mag have similar \textit{peculiarity} regardless of their magnitude. Other clusters in this study do not have many BSSs and they only showed large $P\_F6 - P\_F8$ near limiting magnitude.

\subsection{Individual clusters} \label{sec:cluster_discussion}
\textbf{NGC 2682:}
The detection of stars on the MS in the UV CMDs, suggests that many MS stars have excess UV flux. \citet{Jadhav2019} presented the reasons for UV excess such as the presence of hot WD components, chromospheric activity and hot-spots on contact binaries. Such UV excess detected among the MS stars is unique to NGC 2682; as for all other clusters, only a few stars on the MS show UV excess.
In Fig.~\ref{fig:CMD_2682} (a) and (d), a few WD \texttt{members} and many \texttt{field} stars are found near the WD cooling curve. The CMD location suggests these can be WDs.
Their MPs are low due to astrometric and photometric errors.
As NGC 2682 is a well studied OC, we used deep photometric catalogue of \citet{Williams2018}. They identified hot and faint stars in u,(u $-$ g) plane and carried out spectroscopic observations to confirm the WDs from their atmospheric signatures.
We cross-matched all F148W detections with \citet{Williams2018} catalogue and found ten member WDs and six field WDs, along with the three quasars. Therefore, UVIT observations are well suited to detect WDs in NGC 2682.

\begin{figure} 
\centering
\begin{tabular}{c}
\includegraphics[width=0.45\textwidth]{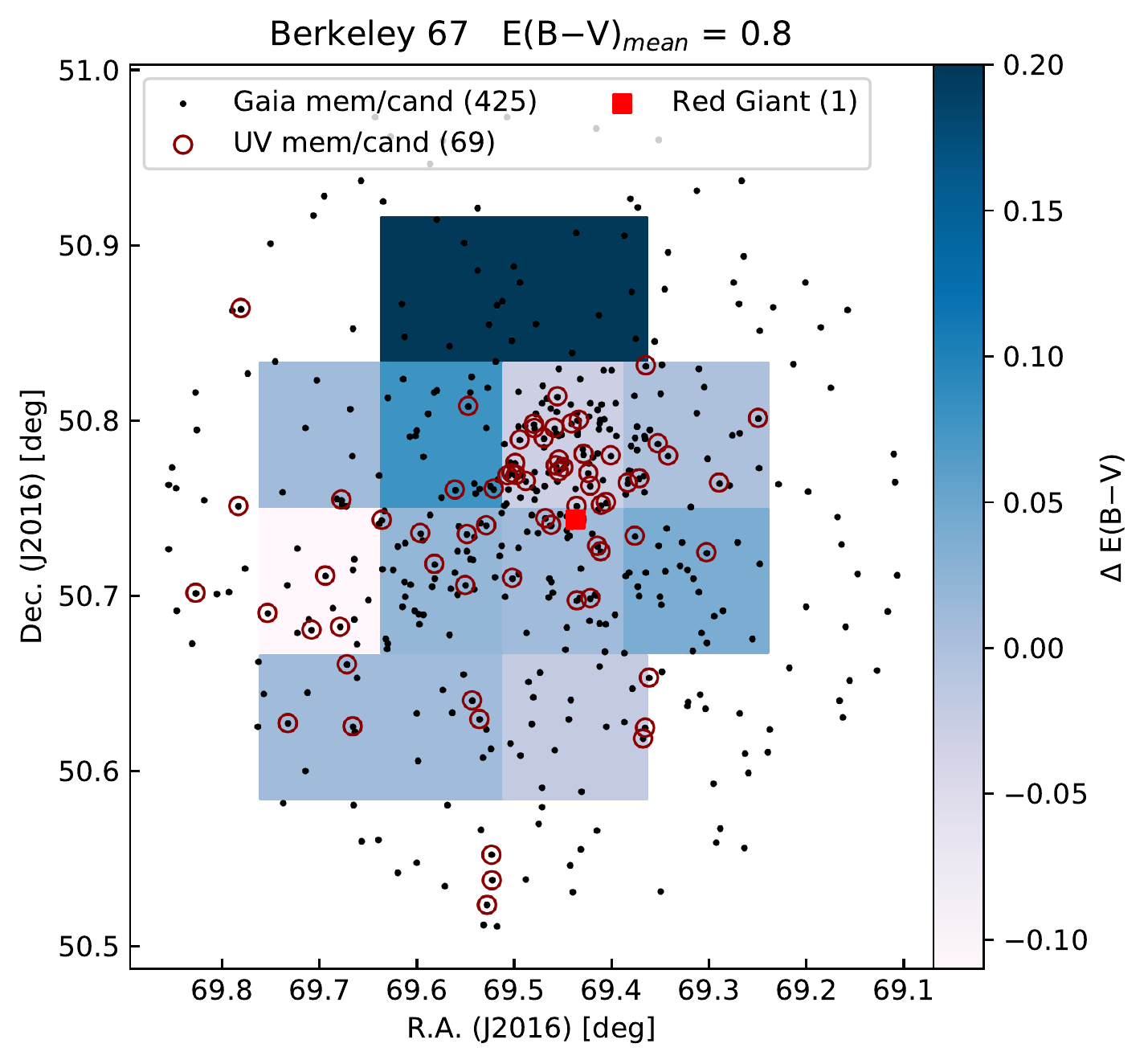}
    \end{tabular}
    \caption{Spatial location of RGs in Berkeley 67 over-plotted on the reddening map. The plot contains all \textit{Gaia} \texttt{members/candidates} (as black dots), N242W \texttt{members/candidates} (as red circles) and the RG detected in N242W filter (as red square).}
    \label{fig:be67_red}
\end{figure}

\textbf{Berkeley 67:}
The cluster has very high reddening (E(B $-$ V) = 0.8 mag). Thus, small relative changes in reddening have a substantial impact on magnitude/colour and cause a broadening of MS in the CMD.  
We tried binning the \textit{Gaia} \texttt{members} spatially and analysed the distributions in the CMD plane. Initial estimates suggested that the E(B $-$ V) values have a range of 0.7 to 1.0 mag. The reddening map is shown in Fig.~\ref{fig:be67_red}. We detected one RG \texttt{member} in NUV which lies in the low reddening region. Further investigation is needed to determine the exact cause of UV brightening of the RG.
Fig.~\ref{fig:CMD_Be67} (c) shows \texttt{members} distributed in MS and sub-giant branch.
From the UV CMD, it indicates the spread at the optical turnoff can be due to subgiant stars or due to differential reddening. As the extinction vector is parallel to the subgiant branch, any differential reddening will increase the spread in the same direction.

\textbf{King 2:}
As the oldest and farthest cluster in this work, only bright BSSs are detected by UVIT. Jadhav (2021; under review) presented the detailed analysis of the detected BSS population, including detection of Extreme Horizontal Branch/subdwarf B type stars as companions to BSSs.

\textbf{NGC 2420:}
We detected the BSS present in the cluster in the UV.
We found $\sim$ 6 stars (out of 59), located on the MS, to show signs of excess UV flux. Some stars are found at the turnoff, and three are \texttt{candidates}. 
One of the \texttt{candidate} has {\sc ruwe} = 3.9 while the other has no \textit{Gaia} colours. High {\sc ruwe} is known to be caused by variability and/or binarity \citep{Deacon2020}. The missing colour/high {\sc ruwe} and excess UV flux points towards a hotter companion or variability. Multi-wavelength analysis and X-ray observations of these stars can shed light on their evolutionary status.

\textbf{NGC 2477:}
We have detected a large number of stars in FUV (108) and NUV (629), hence it is ideal to study the UV properties from the MS up to the red clump. This massive cluster is also ideal to study the UV properties of stars in the broad MS turnoff present in this cluster.
Overall the UV CMDs are aligned with the UV isochrones, not indicating a collective UV brightening alike NGC 2682 among the MS stars. However, there are a couple of stars showing considerable UV excess, which require multi-wavelength study.

\textbf{NGC 6940:}
The UV CMD has some field stars near the WD cooling curve. These can be members or runaway WDs, which are quite faint in {\sc g}.
Some turnoff stars are found to be brighter than the turnoff in the UV CMD, suggesting excess FUV flux.
We detected a giant with \textsc{g = 7.3} mag and \textsc{bp\_rp = 3.7} mag in FUV at the limiting magnitude. It is a variable star of spectral type M5II-III D \citep{Wallerstein1962}. The isochrone suggests that this is likely to be a post-asymptotic giant branch (AGB) star. Due to the low temperature (T$_{eff}^{Gaia}$ = 3355 K), the stellar continuum cannot emit detectable FUV flux. Further study is needed to characterise the emission mechanism.

\subsection{General discussion for all clusters} \label{OC_with_Gaia_UVIT}
\textbf{WD detections:}
Cross-match of \citet{Williams2018} catalogue and \textit{Gaia} EDR3 resulted in only 4 WDs in NGC 2682. While the cross-match with F148W resulted in detection of 16 WDs. The F148W image has detected stars up to 21.7 mag in g-band and 21.8 mag in u-band. This indicates that UV images are more suitable to detect hot WDs as compared to \textit{Gaia}.
CMDs of a few other clusters imply presence of photometric WDs: King 2 (Fig.~\ref{fig:CMD_Be67} (d)), NGC 2477 (Fig.~\ref{fig:CMD_2477} (d)), NGC 6940 (Fig.~\ref{fig:CMD_6940} (b)).
However, comparison with deeper catalogues is required for detecting WDs in these clusters. The membership determination of WDs is challenging due to deficiency of long-baseline deep observations in other open clusters.
Although all the WDs have significant PM errors (\textit{Gaia} EDR3 and \citealt{Yadav2008}), the spread in PM is clearly visible. It is interesting to note that three of the \textit{Gaia} detected WDs in NGC 2682 lie at/just outside the edge of the cluster in the VPD.  

\textbf{Comparison of \textit{Gaia} DR2 and EDR3:}
The membership analysis was done for both \textit{Gaia} DR2 and EDR3. As EDR3 has halved the errors in PM, there were some changes in the members. EDR3 data has led to the addition of sources in the fainter end, which have PM similar to the cluster. We could probe the membership of all the stars in the cluster without any magnitude cutoff due to better accuracy in PM and inclusion of errors in the MP determination. The total percentage of \texttt{candidates} has dropped from 11\% to 5\%, and the VPD distribution of NGC 2682 members was elliptical in DR2 which is now circular in EDR3 data reflecting better handling of systematics in EDR3.

\textbf{Future Improvements:}
There are scopes to improve the membership determination process in future based on the following points:
\begin{enumerate}[leftmargin=*]
    \item The use of RV can constrain the spatial motion of members, however deeper RV data is needed to get cluster membership for fainter stars.
    \item The use of distance from the fiducial isochrone in the CMD could constrain the spread visible in the fainter region of the CMDs. 
    \item We miss some stars with slightly different space velocity, as the primary selection criterion is PM (e.g. 2 PM and RV members from \citealt{Geller2015}). Such stars are important to understand the kinematics of the cluster. Increasing the weightage and accuracy of parallax and CMD location can help identify such stars.
\end{enumerate}

The method developed here is generic and can be applied to non \textit{Gaia} data as well. We recommend a feature-combination similar to F8 (\textsc{ra, dec, pmra, pmdec, pmR0, parallax}/distance, photometric information) to constrain the spread in CMD and VPD. Additionally, comparison with equivalent F6 (Only astrometric information) will be helpful to identify \textit{peculiar} stars in the CMD.

\section{Conclusions and Summary} \label{sec:conclusions}
\begin{enumerate}[leftmargin=*]
    \item We developed an ML-based method to determine the individual stellar membership within OCs using \textit{Gaia} EDR3. We have tried more than 22 different feature-combinations to calculate the MPs. The stars are classified as \texttt{members, candidates} and \texttt{field} using a combination of two {\sc prf} methods. Our primary method (F10) identifies stars which have properties similar to the mean cluster properties and have small systematic errors as \texttt{members}. To incorporate \textit{peculiar} stars (uncommon CMD locations) and stars with large systematic errors, we utilised another method (F6) which only uses spatial location and velocity coordinates. We compared and validated the performance of our methods with past membership studies. Additionally, we created a technique to identify stars with \textit{peculiar} CMD position and demonstrated that it could identify BSSs.
    \item We demonstrated that the {\sc prf} algorithm could be used to determine the MPs in a variety of clusters. It is found to be robust, reproducible, versatile and efficient in various environments (variation in stellar density, reddening, age etc.). We have identified 200--2500 more cluster \texttt{members}, primarily in the fainter MS, compared to previous studies (which used \textit{Gaia} DR2 data). The algorithm presented here is generic and could be changed to suit other data sets or scientific problems. It is editable by selecting different features or creating new features, as required.
    \item We present a catalogue (\textit{Gaia} EDR3 based) of six clusters which provides spatial location, MPs and classification in Table~\ref{tab:cat_Gaia}. The presence of \texttt{candidate} stars suggests a need for better astrometry and photometry, which will be available in future \textit{Gaia} releases and other large scale surveys. We used the \textit{Gaia} catalogue to identify cluster members in UVIT images. We present the UVIT catalogue of six OCs in one or more filters along with its membership information in Table~\ref{tab:cat_UV} (full catalogues are available online). We estimated cluster properties such as mean PM, distance, mean RV and core radii from the identified \texttt{member} population. 
    \item We detected 3 to 700 \texttt{member} stars in various UVIT images of six clusters, apart from $\sim$13\% \texttt{candidates}. We detected BSSs in King 2, NGC 2477, NGC 2420 and NGC 2682. FUV photometry presented here will be used to understand the formation pathways of BSSs. We also detected giant \texttt{members} in FUV (NGC 2682, NGC 6940) and NUV (Berkeley 67, NGC 2477). While most of the NUV detections are expected due to their luminosity and temperature, their FUV detections are unusual. We detected 10 WD members in FUV images of NGC 2682. UV CMDs indicates that there are a few possible WDs in NGC 2477, NGC 2682 and NGC 6940.
    \item As seen in earlier studies, NGC 2682 has unusually high UV bright MS members. We detect no such systematic UV brightening among MS stars in other clusters. Some individual stars do show excess UV flux (RGs, a post-AGB star and a few MS stars). These are good contenders for detailed individual studies. The VPD of NGC 2682 is also notable due to its elliptical shape.
    \item The massive cluster NGC 2477 has 92/576 \texttt{members} detected in FUV/NUV, which will be useful to study the UV properties of stars in the extended turnoff and various evolutionary stages from MS to red clump.
\end{enumerate}

The UV and \textit{Gaia} catalogues provide a comprehensive data set to inter-compare UV emission across various types of clusters and study stellar properties.
We plan to perform a detailed analysis of the interesting sources identified in this study using panchromatic data (such as UVIT, \textit{Gaia} EDR3 and X-ray) in future studies. 

\section*{Data Availability}
The \textit{Gaia} EDR3 data is available at \url{https://gea.esac.esa.int/archive/}. The UVIT images can be accessed through \url{https://astrobrowse.issdc.gov.in/astro_archive/archive/Home.jsp} depending upon their proprietary period.
The membership catalogue and UV photometric catalogues of six clusters are available at CDS via anonymous ftp to \url{cdsarc.u-strasbg.fr}
(130.79.128.5).

\section*{Acknowledgements}

We thank the referee for constructive comments and valuable suggestions. We thank E. Vasiliev for help with \textsc{GaiaTools} and P. Bergeron for providing WD cooling curves for UVIT filters. We thank Deepthi S. Prabhu, Sharmila R. and Samyaday Choudhury for helpful discussions and manuscript preparation. RS thanks the National Academy of Sciences, India (NASI), Prayagraj, for the award of a NASI honorary Scientist position; the Alexander von Humboldt Foundation, Germany for the award of Group linkage long-term research program between IIA, Bengaluru and European Southern Observatory, Munich, Germany, and the Director, IIA for providing institutional, infrastructural support during this work. 
This work was supported by PhD Placement grant, ID 429088188, under the Newton--Bhabha Fund partnership. The grant is funded by the UK Department for Business, Energy and Industrial Strategy and Indian Department of Science and Technology and delivered by the British Council. This work was further supported by the UKRI's (UK Research and Innovation) STFC (Science and Technology Facilities Council) PhD studentship.

\textit{Facilities:} UVIT/\textit{ASTROSAT}, \textit{Gaia}. UVIT project is a result of a collaboration between Indian Institute of Astrophysics (IIA), Bengaluru, The Inter-University Centre for Astronomy and Astrophysic (IUCAA), Pune, Tata Institute of Fundamental Research (TIFR), Mumbai, several centres of Indian Space Research Organisation (ISRO), and Canadian Space Agency (CSA).
This work has made use of data from the European Space Agency (ESA) mission {\it Gaia} (\url{https://www.cosmos.esa.int/gaia}), processed by the {\it Gaia} Data Processing and Analysis Consortium (DPAC, \url{https://www.cosmos.esa.int/web/gaia/dpac/consortium}). 

\textit{Software:} \textsc{GaiaTools} \citep{Vasiliev2019}, {\sc prf} \citep{Reis2019p}, {\sc topcat} \citep{Taylor2005}, {\sc iraf} \citep{Tody1993}, astropy \citep{2013A&A...558A..33A}, matplotlib \citep{Hunter2007}, numpy \citep{oliphant2006guide}, pandas \citep{mckinney-proc-scipy-2010}, scipy \citep{2020SciPy-NMeth}.




\bibliographystyle{mnras}
\bibliography{main}


\appendix

\section{Open clusters under study} \label{sec:intro_1}

\textbf{Berkeley 67} is a $\sim$1 Gyr old OC located at a distance of $\sim$2.45 kpc. It is a low-density cluster with an angular diameter of $\sim$14\arcmin. \citet{Lata2004} carried out deep Johnson UBV and Cousins RI CCD photometry of this cluster while \citet{Maciejewski2007} obtained BV CCD data as part of a survey of 42 open star clusters. Both studies are based on optical CMD of the cluster.

\textbf{King 2} is a $\sim$5 Gyr old OC located at a distance of $\sim$6 kpc towards the Galactic anti-centre direction. It is a faint but rich cluster situated in a dense stellar field. It lags behind the local disc population by 60 to 100 km s$^{-1}$ and could be part of the Monoceros tidal stream \citep{Warren2009}. \citet{Kaluzny1989} obtained BV CCD photometric data for the cluster. A deep Johnson--Cousins UBVR CCD photometric study of the cluster was carried out by \citet{Aparicio1990}. They estimated E(B $-$ V) = 0.31 mag in the direction of the cluster and also indicated the presence of $>25\%$ of binary stars, based on the observed scatter in the CMD of the cluster.

The OC \textbf{NGC 2420} is $\sim$1 Gyr old and located at a distance of $\sim$3 kpc. \citet{Cannon1970} obtained relative PMs and also determined BV photographic magnitudes. The broadband optical CCD photometric study was carried out by \citet{Sharma2006}. The ubyCaH intermediate-band CCD photometry of this star cluster was performed by \citet{Anthony2006}. All these studies indicate that the age of NGC 2420 is older than 1 Gyr.

The intermediate-age ($\sim$0.9 Gyr) southern rich OC \textbf{NGC 2477} is located at a distance of $\sim$1.4 kpc \citep{Hartwick1974, Smith1983, Kassis1997, Eigenbrod2004, Jeffery2011}. This cluster has a metallicity near Solar ([Fe/H] $\sim$ $-$0.17 to 0.07 dex; \citealt{Friel2002, Bragaglia2008}) and a high binary frequency ($\sim$36\%) for the RGs \citep{Eigenbrod2004}. Presence of significant differential reddening (E(B $-$ V) = 0.2 to 0.4 mag) across the cluster was indicated \citep{Hartwick1972, Smith1983, Eigenbrod2004}. Using \textit{Gaia} DR2 data down to $\sim$21 mag, \citet{Gao2018b} identified more than 2000 cluster members. A deep \textit{HST} photometric study of the NGC 2477 was carried out by \citet{Jeffery2011} to identify WD candidates and estimate their age.

\textbf{NGC 2682} (M67) is a nearby OC with an age of $\sim$ 3--4 Gyr \citep{Montgomery1993, Bonatto2015} and located at a distance of $\sim$ 800--900 pc \citep{Stello2016}. It is a well-studied cluster from X-rays to IR \citep{Mathieu1986, Belloni1998, Bertelli2018, Sindhu2018}. There are various studies on the membership determination of NGC 2682 \citep{Sanders1977, Yadav2008, Geller2015, Gao2018c}. It contains stars in various stellar evolutionary phases such as main-sequence (MS), RGs, BSSs, WDs. NGC 2682 contains 38\% photometric binaries \citep{Montgomery1993} and 23\% spectroscopic binaries \citep{Geller2015}. Recently \citet{Sindhu2019} and \citet{Jadhav2019} detected massive and extremely low mass (ELM) WDs with UVIT observations. The presence of 24 BSSs, four yellow stragglers, two sub-subgiants, massive WDs and ELM WDs indicates that constant stellar interactions are happening in NGC 2682.  

\textbf{NGC 6940} is a well-known intermediate-age ($\sim$ 1 Gyr) OC located at a distance of about 0.8 kpc. The membership of the cluster was investigated by \citet{Vasilevskis1957} and \citet{Sanders1972}; while photometric studies were carried out by \citet{Walker1958}, \citet{Johnson1961}, \citet{Larsson1964} and \citet{Jennens1975}. \citet{Baratella2018} presented medium resolution (R $\sim$ 13000), high signal-to-noise (S/N $\sim$ 100), spectroscopic observations of seven RG members.


\section{Supplementary Tables and Figures (Fig.~B3 to Fig.~B11 are only available in the arXiv version)} \label{sec:figures}

\begin{table*}
\centering
\caption{Example of \textit{Gaia} EDR3 membership catalogue 
with MP using {\sc gmm} and ML. The spatial coordinates, {\sc g} and {\sc g\_rp} along with MP obtained by {\sc gmm}, F6, F8 and F10 feature-combinations are included. The `class' column shows the classification according to Eq.~\ref{eq:classif} (M: \texttt{member}, C: \texttt{candidate} and F: \texttt{Field}).}
\label{tab:cat_Gaia}
\begin{tabular}{ccccc ccccc cc}
\toprule
source\_id &         RAdeg &        DEdeg &  g\_mag &  g\_rp & qf &   P\_F6 &   P\_F8 &  P\_F10 &  P\_GMM & class &  cluster \\
\midrule
 260364731415812736 &  69.623768 &  50.538089 &  19.93 &  1.27 &  0 &  0.431 &  0.345 &  0.038 &     --- &     F &  Berkeley\_67 \\
 260364804431166080 &  69.683631 &  50.556330 &  20.43 &  1.06 &  0 &  0.342 &  0.192 &  0.039 &     --- &     F &  Berkeley\_67 \\
 260364804433635840 &  69.687125 &  50.552245 &  19.76 &  1.10 &  1 &  0.081 &  0.092 &  0.104 &     --- &     F &  Berkeley\_67 \\
 260364834495034880 &  69.653163 &  50.556060 &  19.95 &  1.06 &  1 &  0.370 &  0.358 &  0.532 &     --- &     F &  Berkeley\_67 \\
 260364838790438784 &  69.650741 &  50.557053 &  20.07 &  1.04 &  1 &  0.118 &  0.133 &  0.136 &     --- &     F &  Berkeley\_67 \\
\bottomrule
\end{tabular}
\end{table*} 

\begin{table*}
\centering
\caption{Example of photometric catalogue of all the detected stars in the UVIT images of NGC 6940. The catalogue includes UV magnitudes and errors along with the membership classification. Similar tables for each cluster are available online. The magnitudes of saturated stars are listed as `F169M\_sat'. The last column shows the classification according to Eq.~\ref{eq:classif} (M: \texttt{member}, C: \texttt{candidate} and F: \texttt{Field}).}
\label{tab:cat_UV}
\begin{tabular}{ccc ccc cc}
\toprule
    RAdeg &     DEdeg &  F169M &  F169M\_sat &  e\_F169M &  P\_F10 &   P\_F6 & class \\
\midrule
 308.6433 &  28.25829 &  19.78 &        --- &     0.08 &  0.003 &  0.005 &     F \\
 308.7407 &  28.22939 &  19.88 &        --- &     0.09 &    --- &    --- &   --- \\
 308.6312 &  28.23331 &  19.70 &        --- &     0.10 &  0.960 &  0.992 &     M \\
 308.9526 &  28.28288 &  17.86 &        --- &     0.04 &  0.732 &  0.006 &     C \\
 308.8039 &  28.35747 &  20.92 &        --- &     0.18 &  0.972 &  0.994 &     M \\
\bottomrule
\end{tabular}

\end{table*} 

\begin{figure*}
\centering
\includegraphics[width=\textwidth]{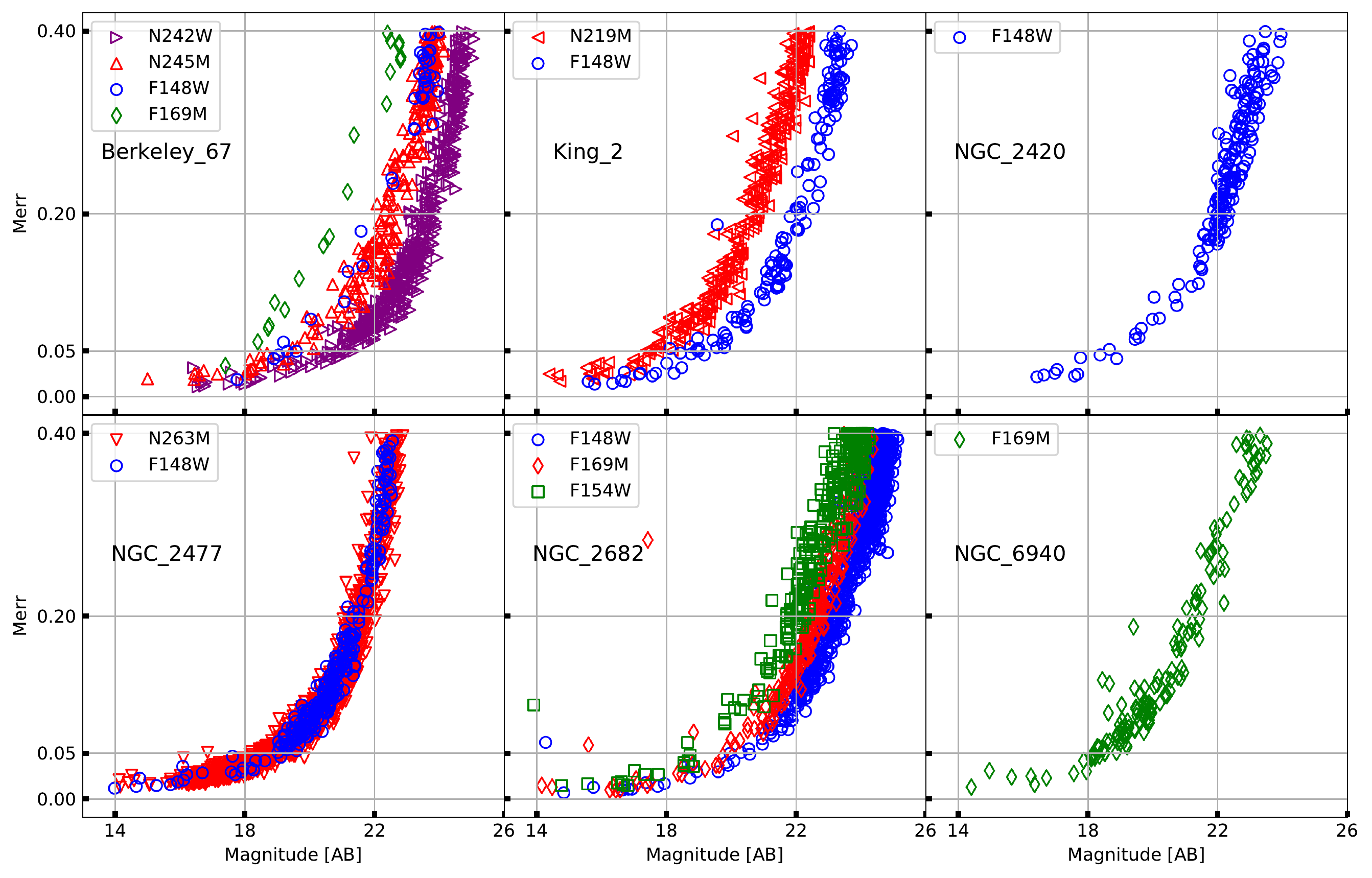}
    \caption{The photometric error in the magnitudes. Each subplot shows the magnitude-error plots for a cluster in all available filters. Please refer to the electronic version of the journal for the colour version.}
    \label{fig:mag_err}
\end{figure*} 
\onecolumn
\begin{landscape}
\begin{figure*}
   \centering
    \includegraphics[height=0.8\textwidth]{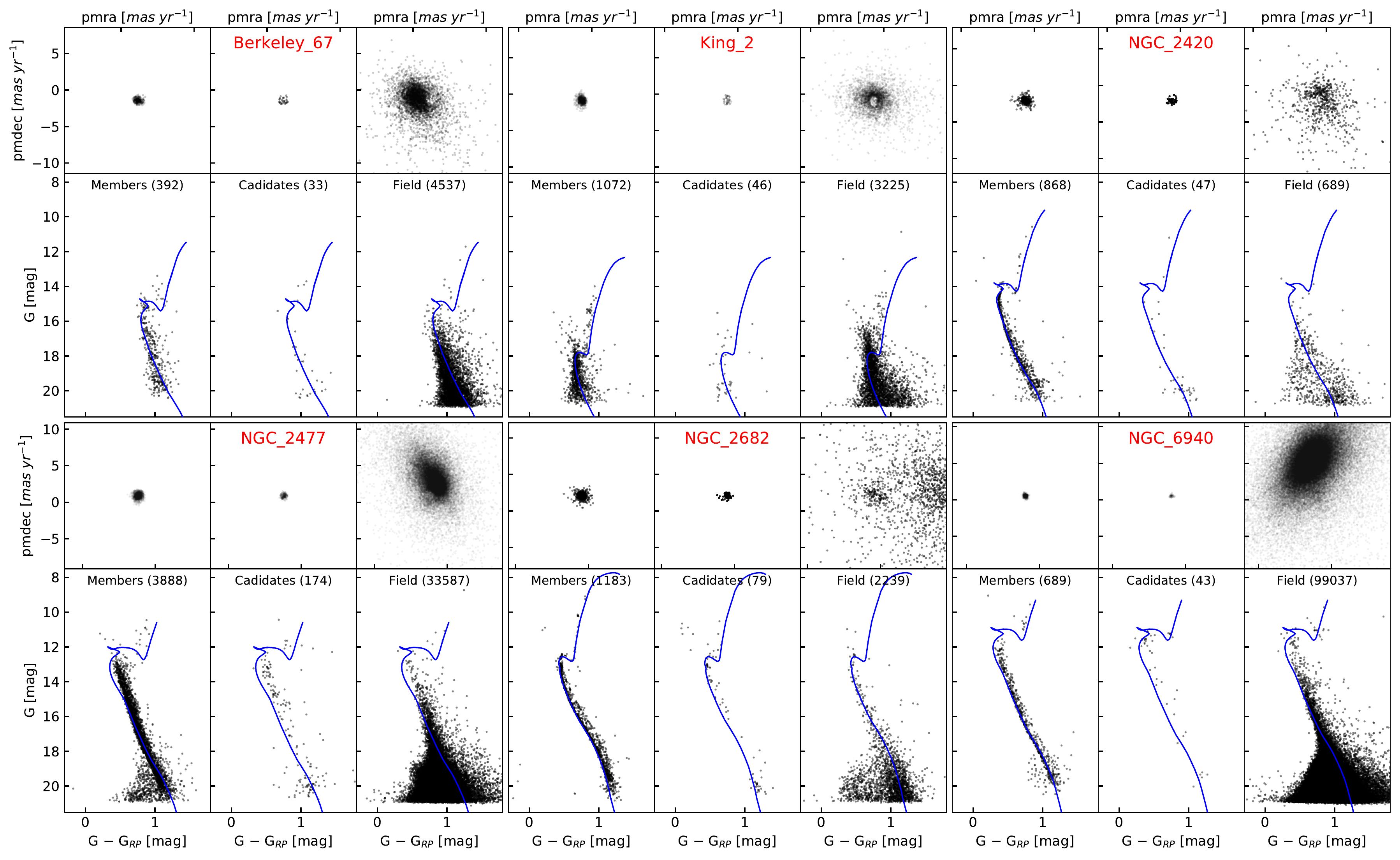}
    \caption{VPDs and CMDs of the six clusters. First and third rows are VPDs of \texttt{members, candidates} and \texttt{field} for respective clusters. 
    Second and fourth rows are CMDs of \texttt{members, candidates} and \texttt{field} for respective clusters.
    Isochrones are plotted as blue line for reference.}
    \label{fig:CV_combined}
\end{figure*}

\end{landscape}

\begin{figure*}
   \centering
    \includegraphics[width=0.93\textwidth]{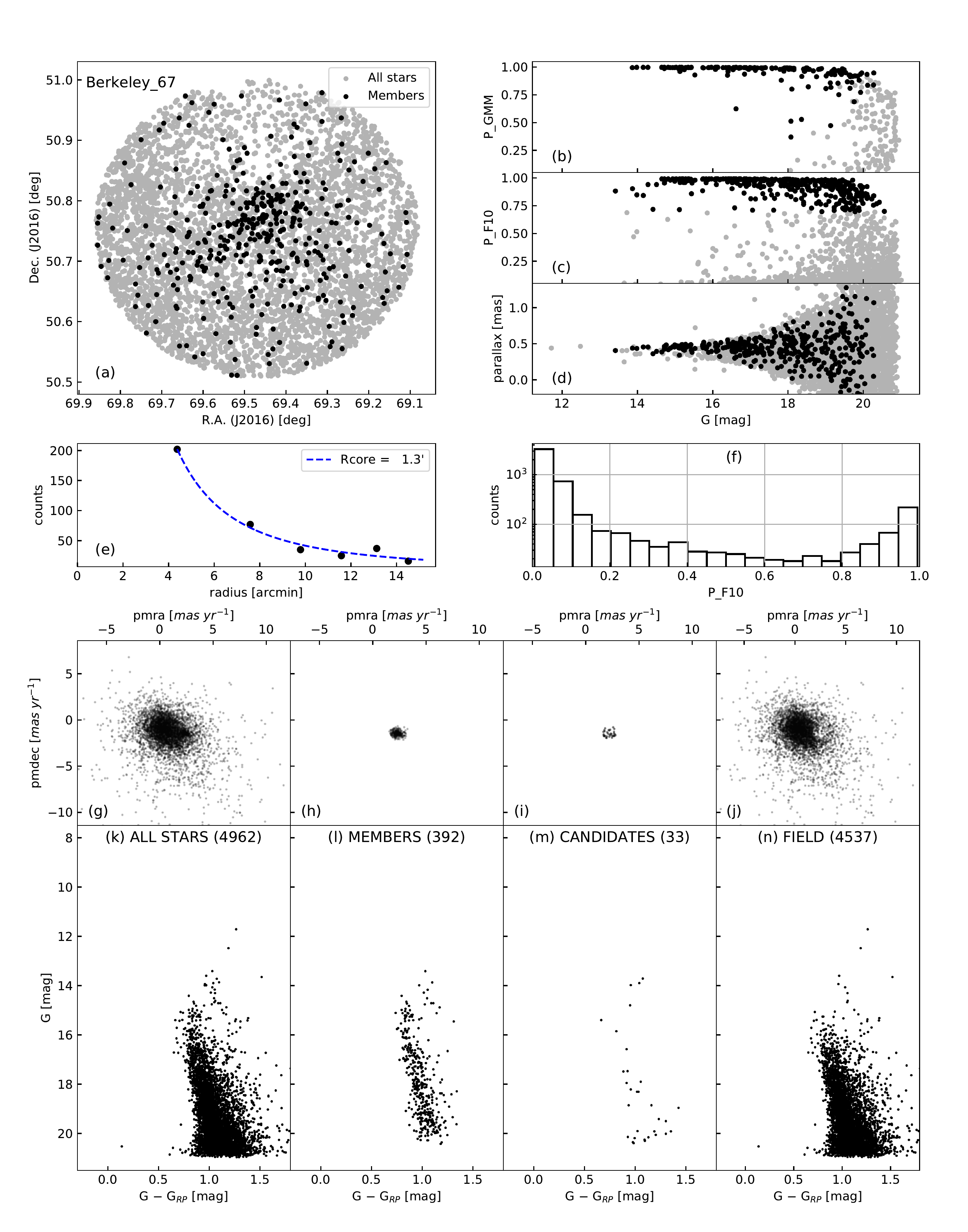}
    \caption{(a) Spatial distribution for Berkeley 67. All \textit{Gaia} stars are denoted by grey dots and cluster \texttt{members} are denoted by black points. (b)--(d) Distribution of {\sc g} magnitude with P\_GMM, P\_F10.3 and parallax. (e) Calculation of core radius by fitting radial density with King's surface density profile. (f) Histogram of P\_10.3. The bimodal Moffat profile is shows with dashed curve, while the minima of the profile is represented by dot-dashed line. (g)--(j) and (k)--(n) show the VPD and CMD of all stars, \texttt{members, candidates} and \texttt{field} respectively. (k)--(n) show the number of stars plotted in the brackets along with the Padova isochrone as a blue curve.}
    \label{fig:rVC_be67}
\end{figure*}

\begin{figure*}
   \centering
    \includegraphics[width=0.97\textwidth]{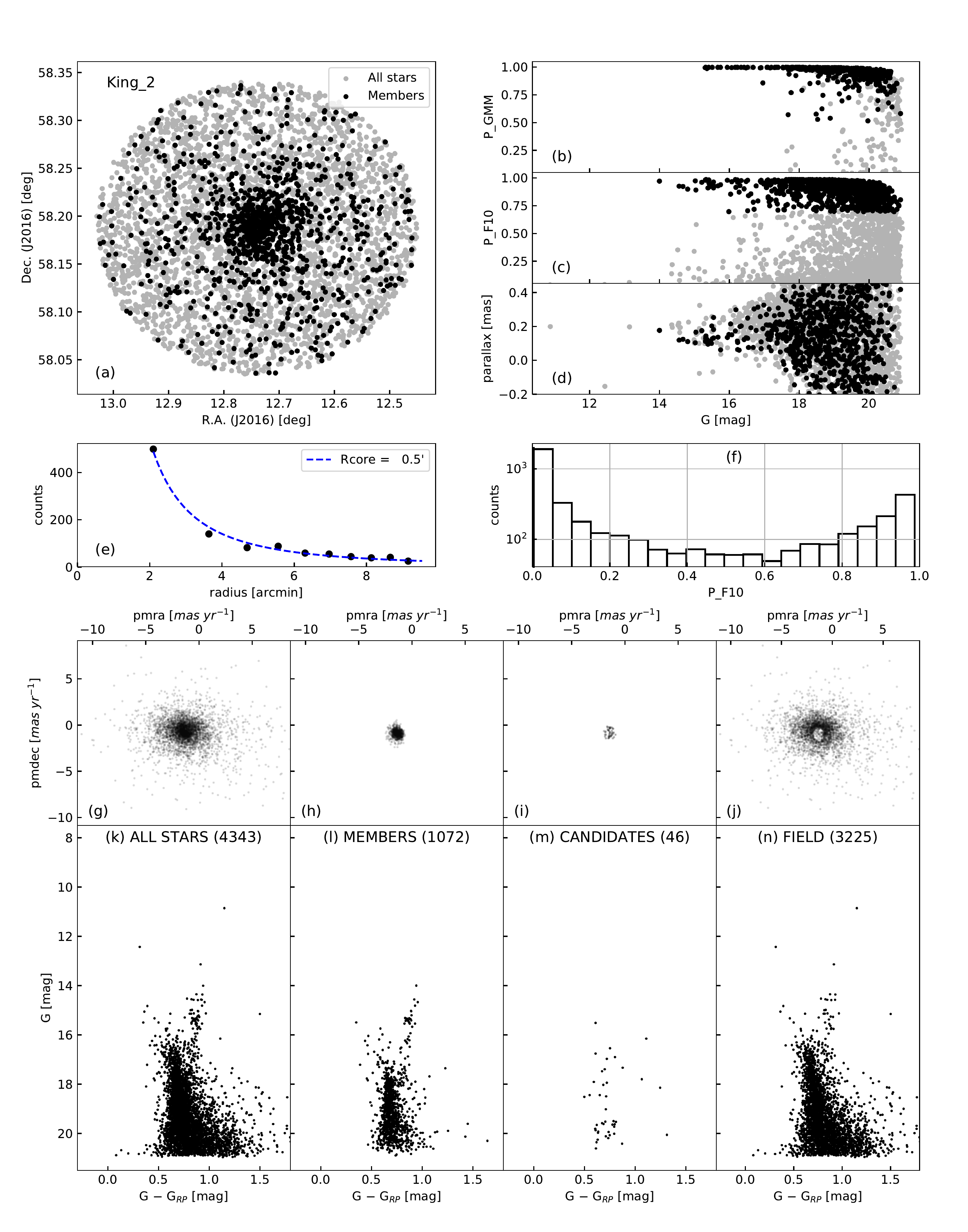}
    \caption{Spatial distribution, VPD and CMDs of King 2. The subplot descriptions are same as Fig.~\ref{fig:rVC_be67}.}
    \label{fig:rVC_K2}
\end{figure*}
\begin{figure*}
   \centering
    \includegraphics[width=0.97\textwidth]{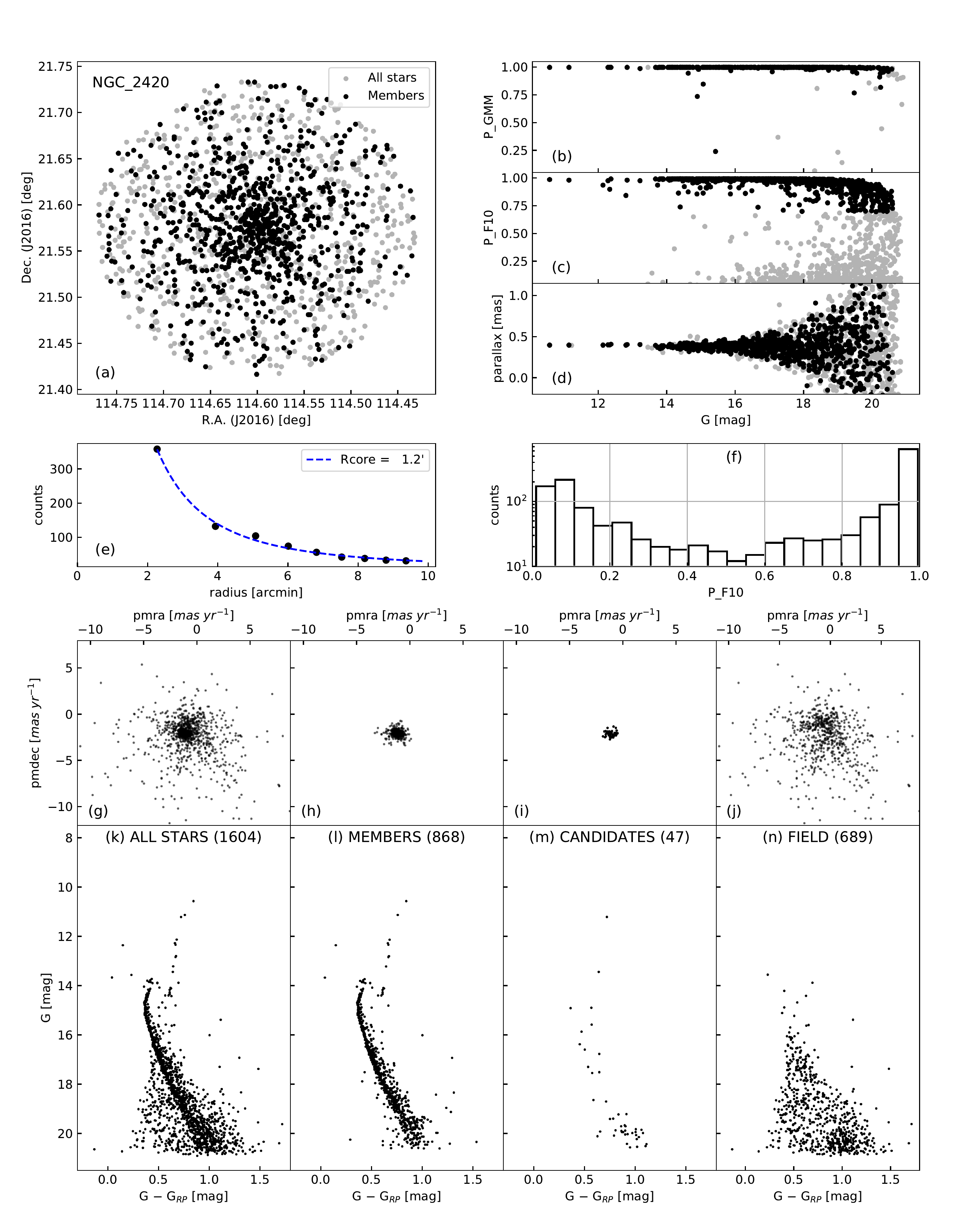}
    \caption{Spatial distribution, VPD and CMDs of NGC 2420. The subplot descriptions are same as Fig.~\ref{fig:rVC_be67}.}
    \label{fig:rVC_2420}
\end{figure*}
\begin{figure*}
   \centering
    \includegraphics[width=0.97\textwidth]{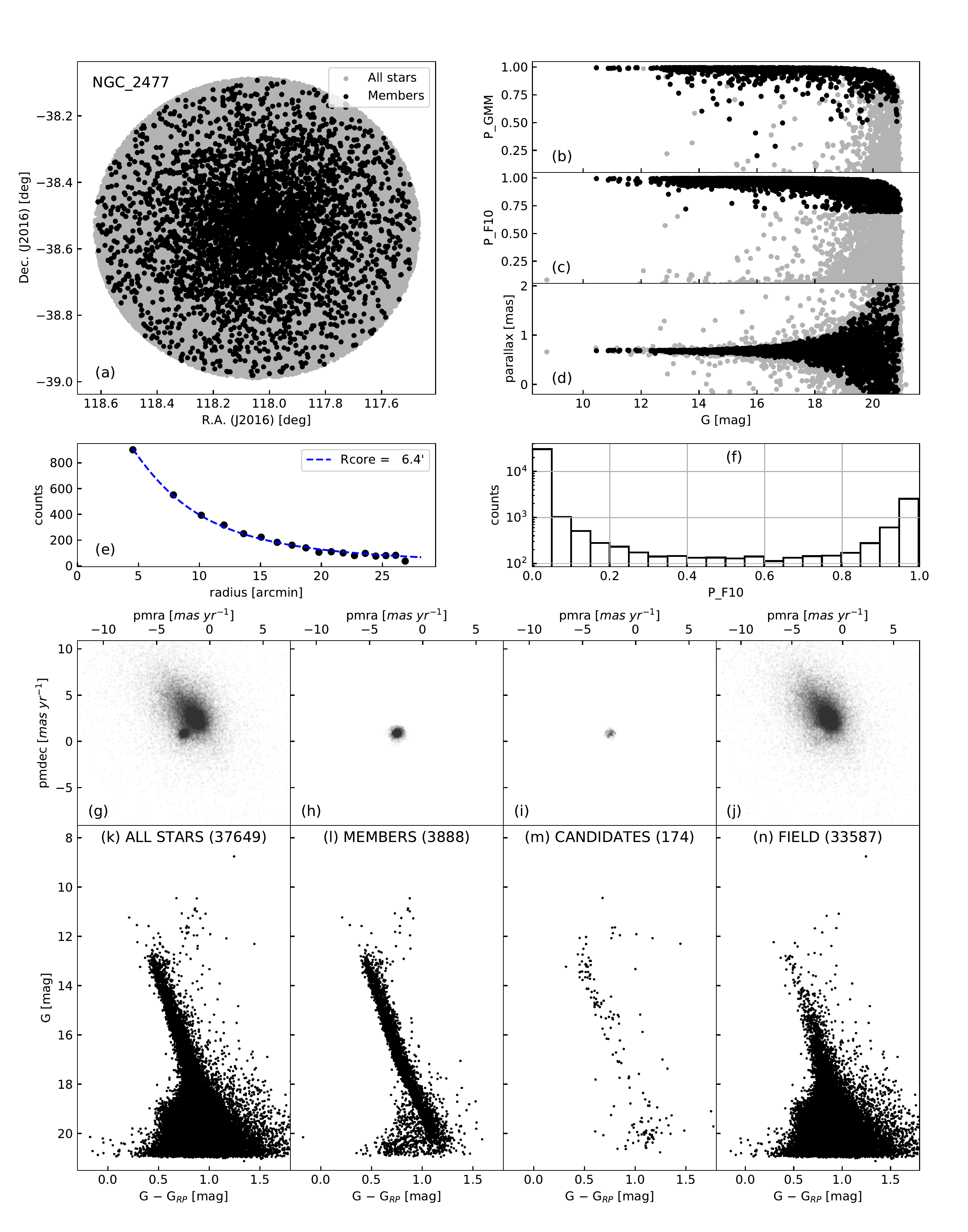}
    \caption{Spatial distribution, VPD and CMDs of NGC 2477. The subplot descriptions are same as Fig.~\ref{fig:rVC_be67}.}
    \label{fig:rVC_2477}
\end{figure*}
\begin{figure*}
   \centering
    \includegraphics[width=0.97\textwidth]{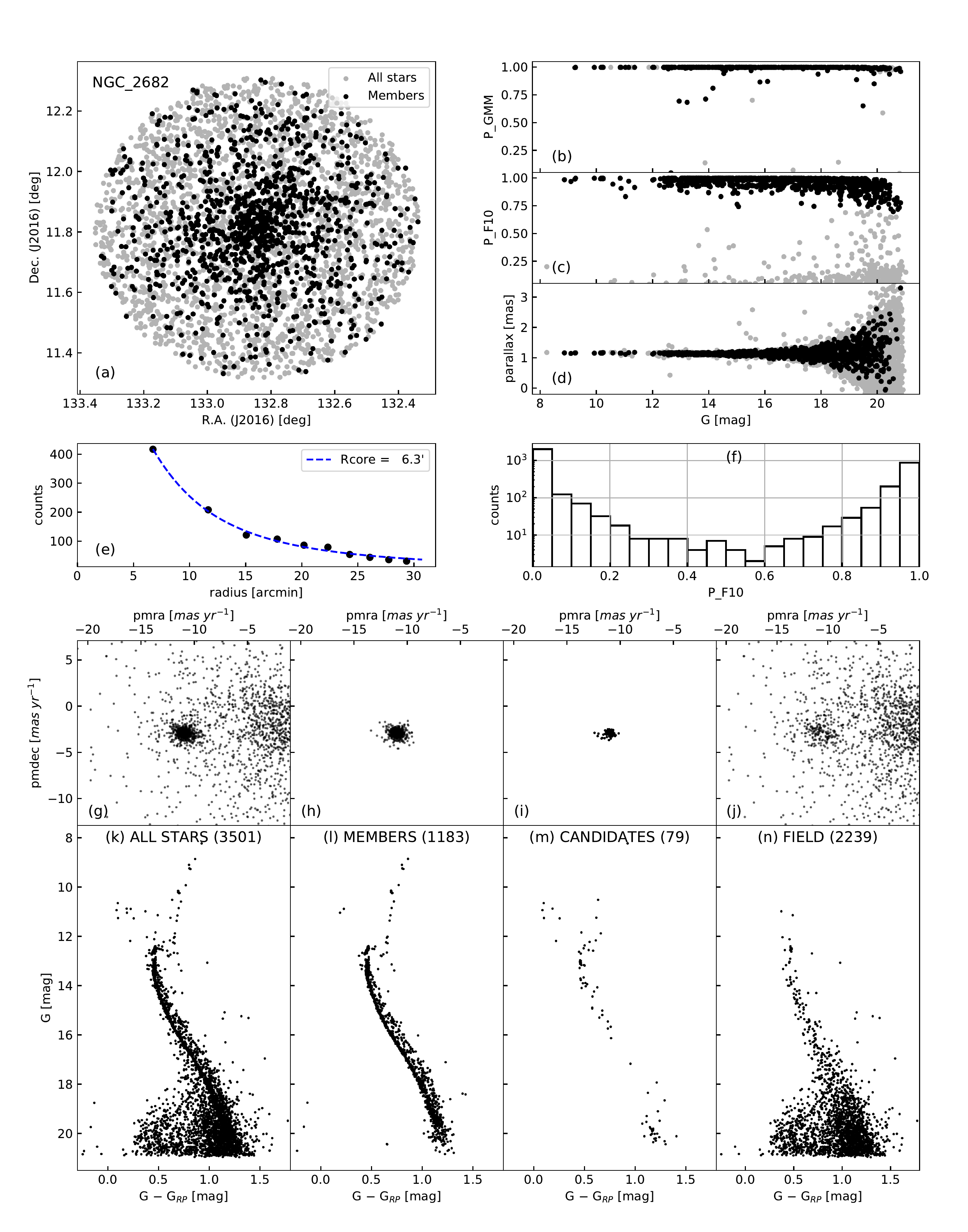}
    \caption{Spatial distribution, VPD and CMDs of NGC 2682. The subplot descriptions are same as Fig.~\ref{fig:rVC_be67}.}
    \label{fig:rVC_2682}
\end{figure*}
\begin{figure*}
   \centering
    \includegraphics[width=0.97\textwidth]{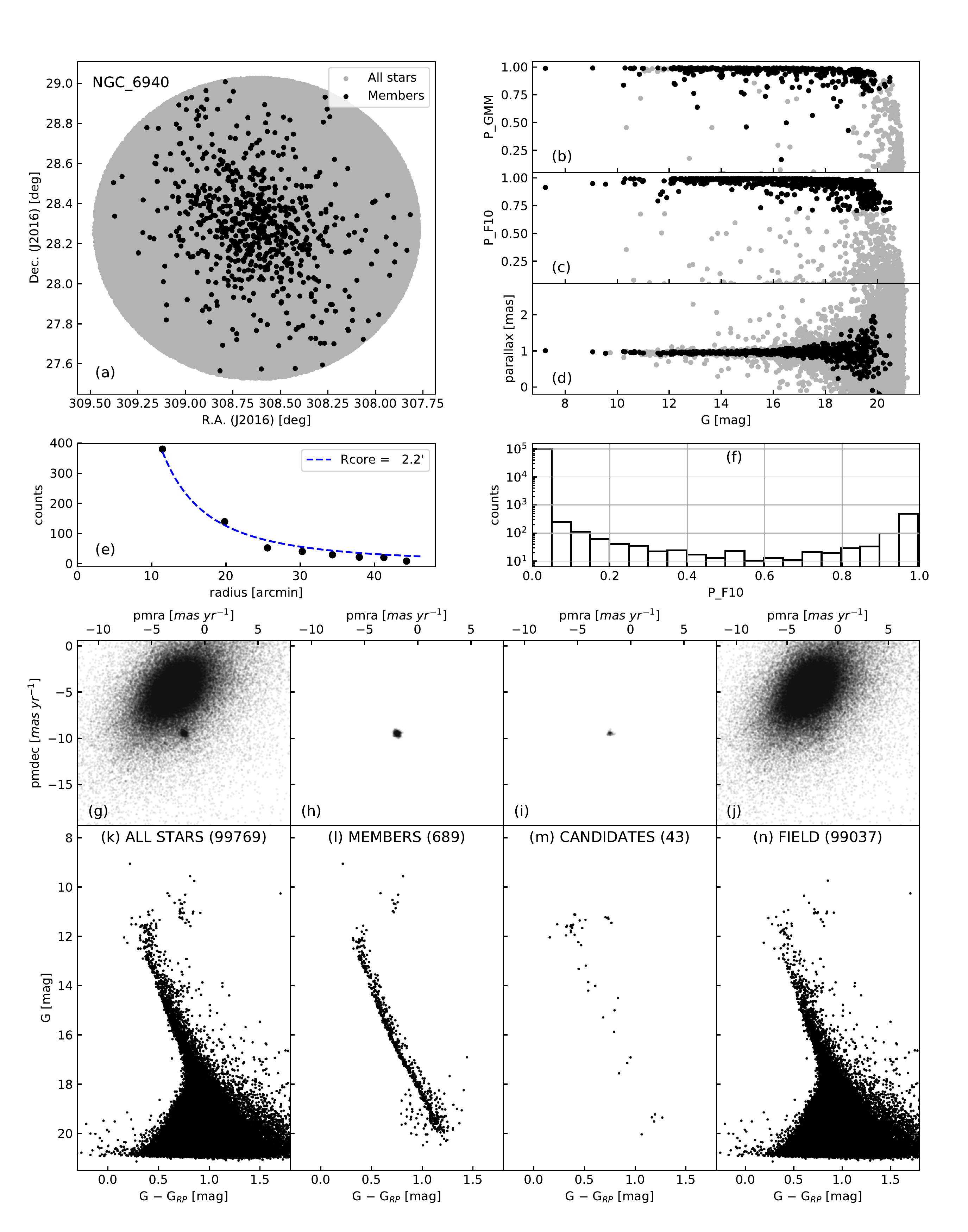}
    \caption{Spatial distribution, VPD and CMDs of NGC 6940. The subplot descriptions are same as Fig.~\ref{fig:rVC_be67}.}
    \label{fig:rVC_6940}
\end{figure*}

\begin{landscape}

\begin{figure*}
  \centering
    \includegraphics[height=0.89\textwidth]{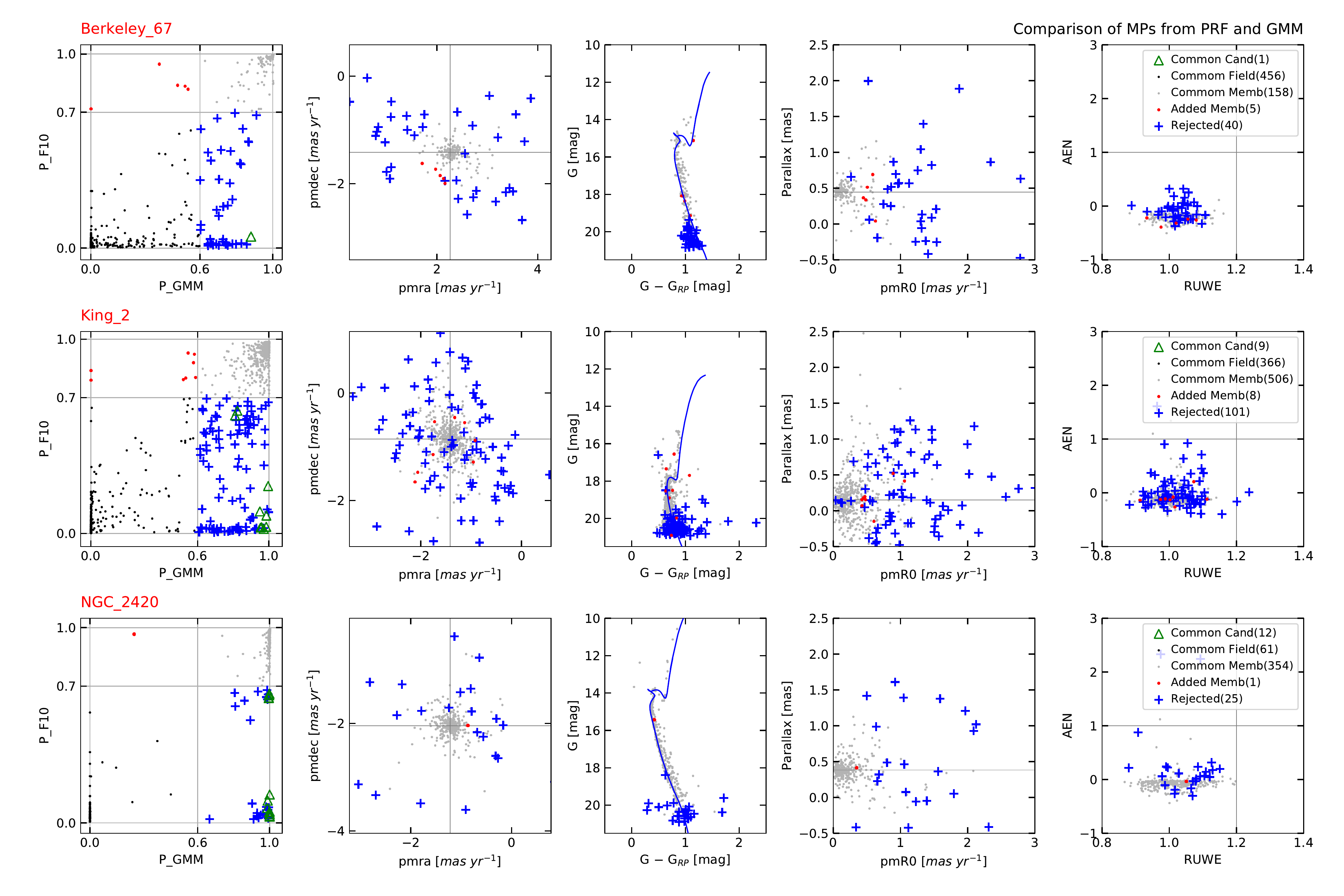}
    \caption{Comparison of MPs from {\sc prf} and {\sc gmm}. Common members (P\_F10.3 $>$ cutoff {\sc and} P\_GMM) $>$ 0.6) are shown as grey dots, common candidates (P\_F10.3 $<$ cutoff {\sc and} P\_F6.4 $>$ cutoff {\sc and} P\_GMM) $>$ 0.6) are shown as green triangles, common field stars (P\_F10.3 $<$ cutoff {\sc and} P\_F6.4 $<$ cutoff {\sc and} P\_GMM) $<$ 0.6) are shown as black dots, added members (P\_F10.3 $>$ cutoff {\sc and} P\_GMM) $<$ 0.6) are shown as red dots and rejected stars (P\_F10.3 $<$ cutoff {\sc and} P\_GMM) $>$ 0.6).}
    \label{fig:GMM_comp_0}
\end{figure*}
\begin{figure*}
  \centering
    \includegraphics[height=0.9\textwidth]{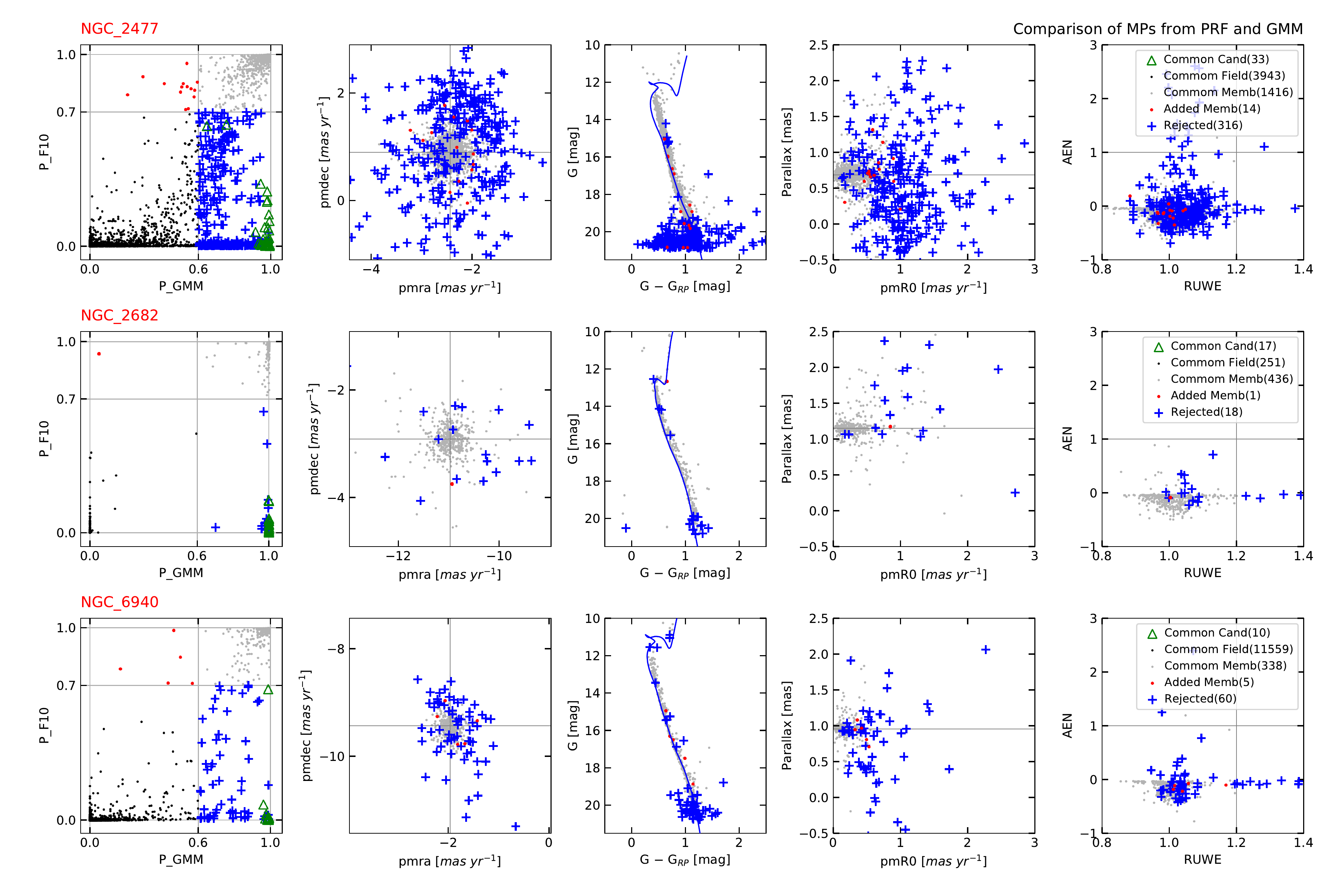}
    \contcaption{}
\end{figure*}

\begin{figure*}
  \centering
    \includegraphics[height=0.9\textwidth]{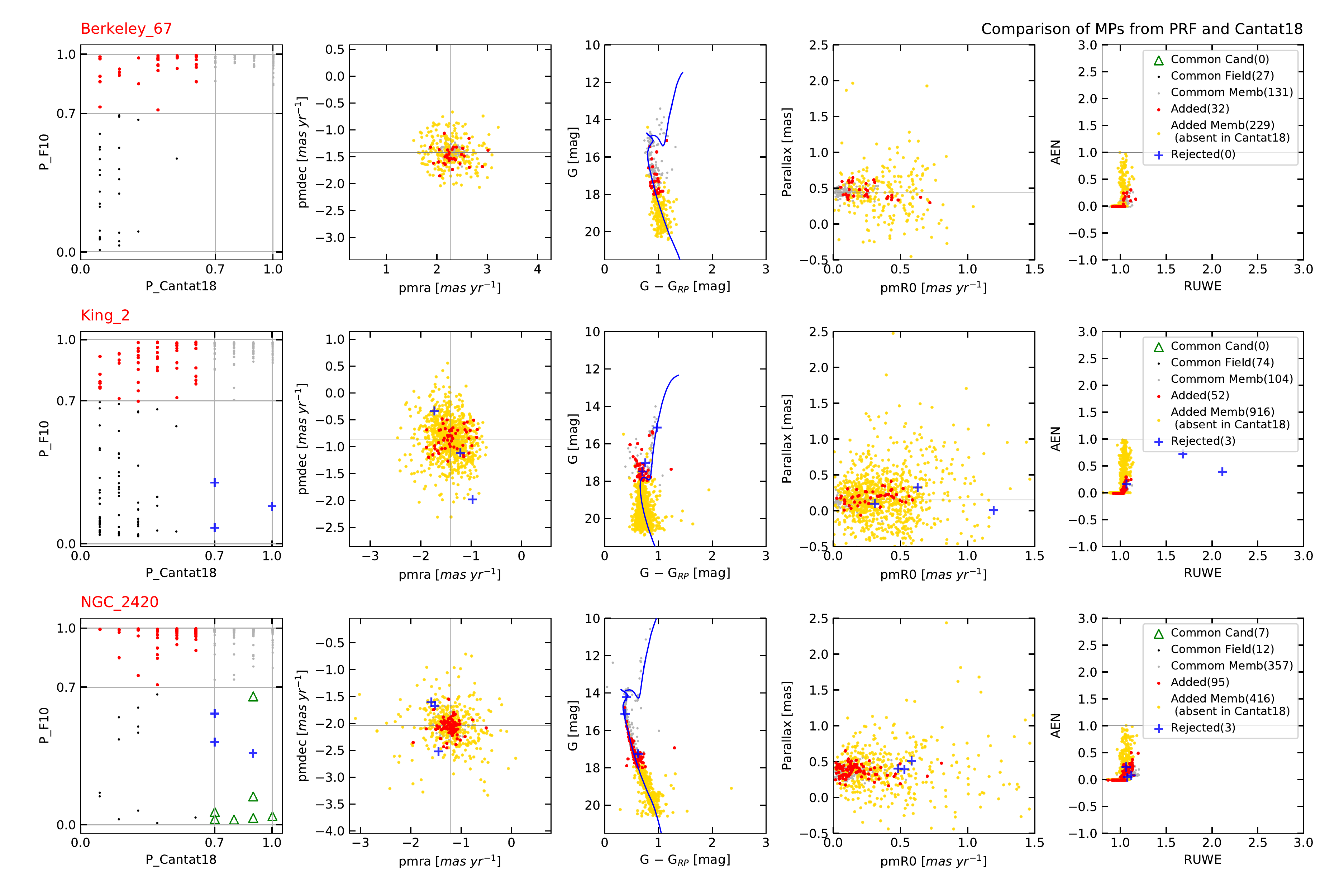}
    \caption{Comparison of MPs from {\sc prf} and \citet{Cantat2018}. The markers for different types of stars and the individual panels are similar to Fig.~\ref{fig:GMM_comp_0}.}
    \label{fig:Cant18_comp_0}
\end{figure*}
\begin{figure*}
  \centering
    \includegraphics[height=0.9\textwidth]{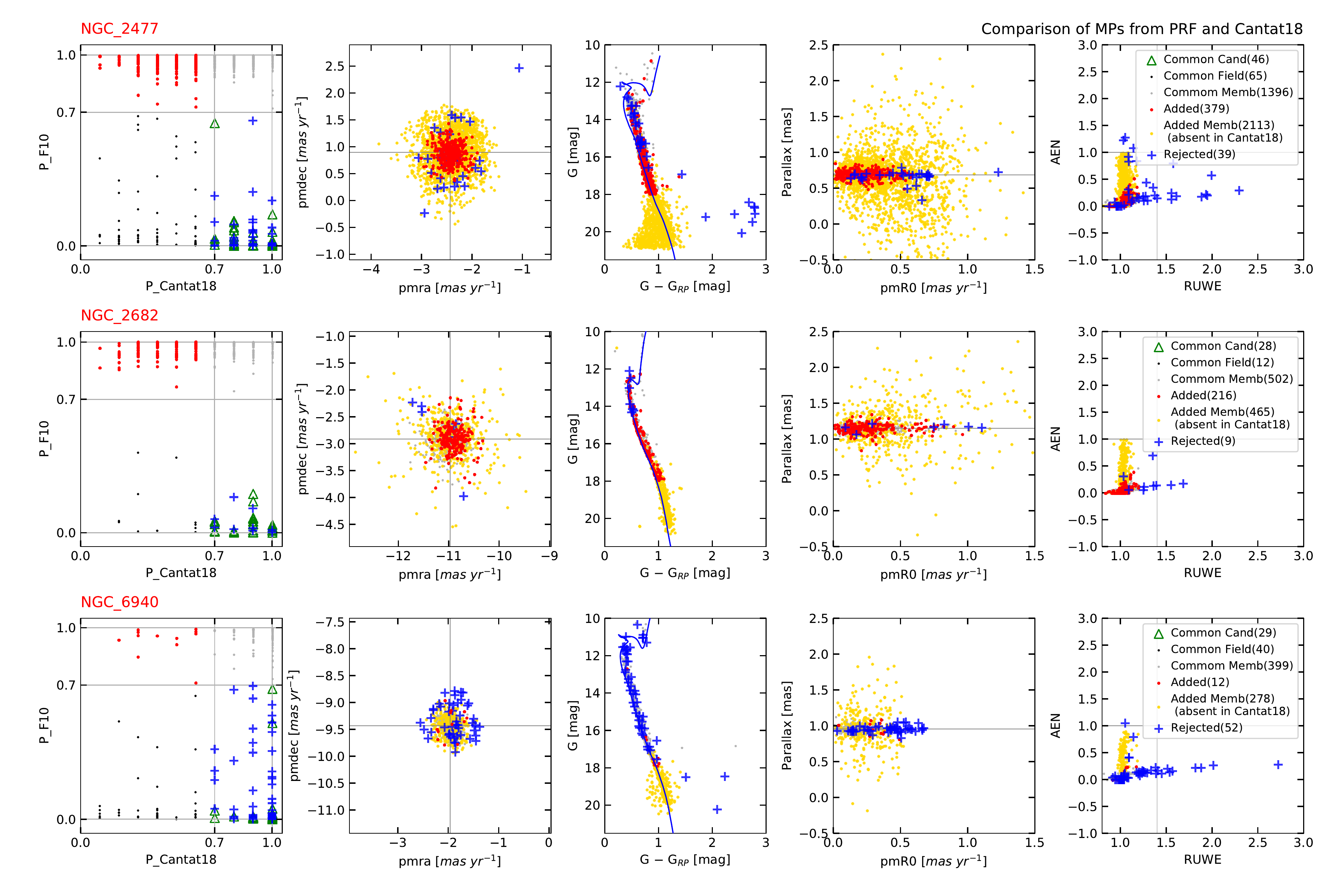}
    \contcaption{}
\end{figure*}

\begin{figure}
  \centering
    \includegraphics[height=0.9\textwidth]{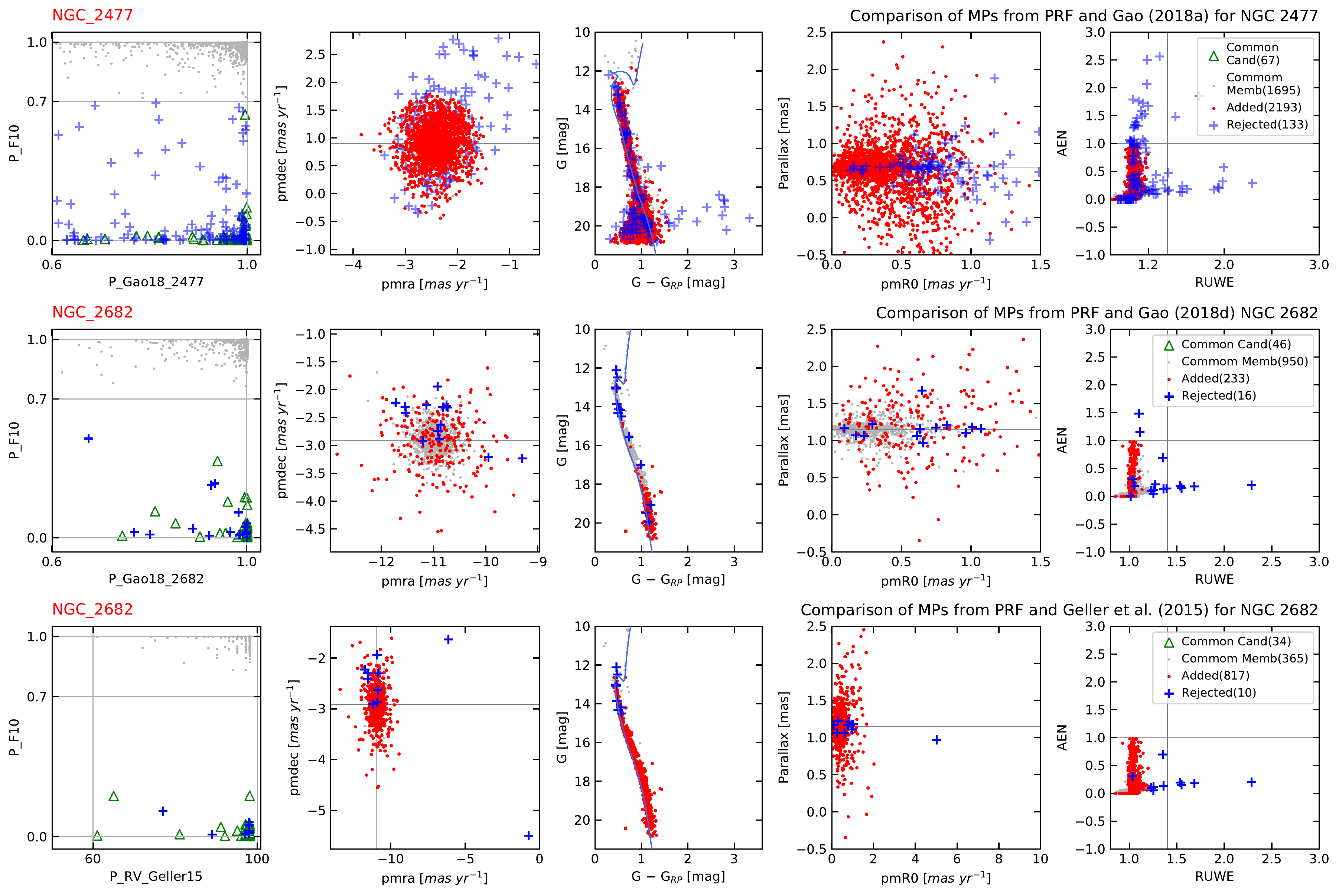}
    \caption{Analysis of membership from {\sc prf} and \citet{Gao2018b,Gao2018c} and \citet{Geller2015}. The markers for different types of stars and the individual panels are similar to Fig.~\ref{fig:GMM_comp_0}.}
    \label{fig:gao_comp}
\end{figure}

\end{landscape}

\bsp	
\label{lastpage}
\end{document}